\documentclass[a4paper,11pt,amsmath,amssymb]{article} 
\usepackage{jheppub}
\usepackage{amsmath} 
\usepackage{array,relsize,float}
\usepackage{pstricks}
\usepackage{color} 
\usepackage{amssymb}
\usepackage{slashed}
\usepackage{tikz-cd}
\usepackage{graphicx} 
\usepackage{epsfig}
\usepackage{hyperref}
\usepackage{pdfpages}
\usepackage{array}
\usepackage{tikz-cd}
\usepackage{amsfonts,layout,appendix,subfigure}
\allowdisplaybreaks
\input{epsf} 

\def\qon#1{q_{#1,0}^{(+)}}

\renewcommand{\thefigure}{\arabic{figure}}

\def\beq{\begin{equation}} \def\eeq{\end{equation}}
\def\beqn{\begin{eqnarray}} \def\eeqn{\end{eqnarray}}
 \def\to{\rightarrow}
\def\nn{\nonumber}

\def\Eq#1{Eq.~(\ref{#1})}

\def\beq{\begin{equation}}
\def\eeq{\end{equation}}
\def\bea{\begin{eqnarray}}
\def\eea{\end{eqnarray}}
\def\beqn{\begin{eqnarray}} \def\eeqn{\end{eqnarray}}
\def\beeq{\begin{eqnarray}}
\def\eeeq{\end{eqnarray}}

\def\nn{\nonumber}
\def\Eq#1{Eq.~(\ref{#1})}

\def\qb{\mathbf{q}}

\def\ii{\imath 0}

\pdfoutput=1

\newcommand{\valencia}{Instituto de F\'{\i}sica Corpuscular, Universitat de Val\`{e}ncia -- Consejo Superior de Investigaciones Cient\'{\i}ficas, Parc Cient\'{\i}fic, E-46980 Paterna, Valencia, Spain.}
\newcommand{\culiacana}{Facultad de Ciencias F\'isico-Matematicas, Universidad Aut\'onoma de Sinaloa, Ciudad Universitaria, CP 80000 Culiac\'an, Mexico.}
\newcommand{\culiacanb}{Facultad de Ciencias de la Tierra y el Espacio, Universidad Aut\'onoma de Sinaloa, Ciudad Universitaria, CP 80000 Culiac\'an, Mexico.}

\begin{document}

\title{Universal opening of four-loop scattering amplitudes to trees}
\author[a,b,c]{Selomit Ram\'irez-Uribe,}
\author[b]{Roger J. Hern\'andez-Pinto,}
\author[a]{Germ\'an Rodrigo,}
\author[a]{German~F. R.~Sborlini} 
\author[a]{and William~J.~Torres~Bobadilla}
\affiliation[a]{\valencia}
\affiliation[b]{\culiacana}
\affiliation[c]{\culiacanb}

\emailAdd{norma.selomit.ramirez@ific.uv.es}
\emailAdd{roger@uas.edu.mx}
\emailAdd{german.rodrigo@csic.es}
\emailAdd{german.sborlini@ific.uv.es}
\emailAdd{william.torres@ific.uv.es}

\preprint{IFIC/20-29}

\date{February 8, 2020}


\abstract{
The perturbative approach to quantum field theories has made it possible to obtain incredibly accurate theoretical predictions in 
high-energy physics.  Although various techniques have been developed to boost the efficiency of these calculations, 
some ingredients remain specially challenging. This is the case of multiloop scattering amplitudes that constitute a hard bottleneck to solve. 
In this paper, we delve into the application of a disruptive technique based on the loop-tree duality theorem,
which is aimed at an efficient computation of such objects by opening the loops to nondisjoint trees.
We study the multiloop topologies that first appear at four loops and assemble them in a clever and general expression, 
the N$^4$MLT {\it universal topology}. This general expression enables to open any scattering amplitude of up to four loops, 
and also describes a subset of higher order configurations to all orders.
These results confirm the conjecture of a factorized opening in terms of simpler known subtopologies, 
which also determines how the causal structure of the entire loop amplitude is characterized by the causal structure 
of its subtopologies. In addition, we confirm that the loop-tree duality representation of the N$^4$MLT universal topology
is manifestly free of noncausal thresholds, thus pointing towards a remarkably more stable numerical 
implementation of multiloop scattering amplitudes.
}

\maketitle


\section{Introduction}

The impressive progress in the understanding of the fundamental building blocks of Nature is due to the ability to extract theoretical predictions 
from Quantum Field Theories. The perturbative framework has proven to be extremely efficient for that purpose, nevertheless, the continuous 
effort to reach better predictions has revealed critical challenges. The main bottleneck to automate higher perturbative orders is the study 
of vacuum quantum fluctuations associated to Feynman loop diagrams. These mathematical objects exhibit a complex
behaviour of physical and unphysical singularities, which prevents straightforward numerical calculations.  
Likewise, the high luminosity achieved by collider machines such as the CERN's LHC~\cite{Mangano:2020icy} 
and future colliders~\cite{Abada:2019lih,Abada:2019zxq,Abada:2019ono,Benedikt:2018csr,Bambade:2019fyw,Djouadi:2007ik,Roloff:2018dqu,CEPCStudyGroup:2018ghi} 
is pushing the precision frontier towards even more accurate theoretical predictions and better understanding of the behaviour of such quantum objects. 

Nowadays, predictions ranging from next-to-leading to even next-to-next-to-next-to leading order have been calculated for several processes
of interest at high energy colliders~\cite{Anastasiou:2015vya,Mondini:2019gid,Mistlberger:2018etf,Ahmed:2016otz,Bonetti:2017ovy,Dreyer:2016oyx,Cieri:2018oms,Cieri:2020ikq}.
Since the numerical evaluation of integrals at multi-loop level requires a careful treatment of singularities, new methods need to be proposed to achieve better theoretical predictions.

The loop-tree duality (LTD)~\cite{Catani:2008xa,Bierenbaum:2010cy,Bierenbaum:2012th,Tomboulis:2017rvd,Runkel:2019yrs,Capatti:2019ypt,Verdugo:2020kzh} features a manifest distinction between physical and unphysical singularities
at integrand level~\cite{Buchta:2014dfa,Aguilera-Verdugo:2019kbz}, opening an alternative framework to perform more efficient calculations. 
This knowledge was crucial for developing the four dimensional unsubtraction (FDU)~\cite{Hernandez-Pinto:2015ysa,Sborlini:2016gbr,Sborlini:2016hat,Driencourt-Mangin:2019sfl}, which allows to combine real and virtual corrections into a single numerically-stable integral. 
As other methods proposed in the literature~\cite{Soper:1999xk,Fazio:2014xea,Freedman:1991tk,Battistel:1998sz,Wu:2002xa,Pittau:2012zd,Gnendiger:2017pys,Pozzorini:2020hkx,Cherchiglia:2020iug}, FDU is aimed at performing most of the calculations directly in the four physical
dimensions of the space-time. 
Additionally, the LTD formalism posses others features that convert it into a promising technique for tackling higher-order computations. 
For instance, the number of integration variables in numerical implementations is independent of the number of external legs~\cite{Buchta:2015wna,Buchta:2015xda,Driencourt-Mangin:2019yhu,Capatti:2019edf,Jurado:2017xut}.
On top of that, LTD efficiently provides asymptotic expansions~\cite{Beneke:1997zp}, \cite{Driencourt-Mangin:2017gop,Plenter:2019jyj,Plenter:2020lop},
and it also constitutes a promising strategy towards local renormalization approaches~\cite{Driencourt-Mangin:2019aix}.

It was recently conjectured in Ref.~\cite{Verdugo:2020kzh} that LTD straightforwardly leads to extremely compact and manifestly causal representations of scattering amplitudes to all orders. This pattern was explicitly proven
for a series of multiloop topologies, the maximal loop topology (MLT), next-to-maximal (NMLT) and next-to-next-to-maximal (N$^2$MLT)
that are characterized by $L+1$, $L+2$ and $L+3$ sets of propagators, respectively, with each set categorized by the dependence 
on a specific loop momentum or a linear combination of the $L$ independent loop momenta.
Remarkably, their analytic dual representations are inherently free of unphysical singularities, 
and the causal structure can be interpreted in terms of entangled causal thresholds~\cite{Aguilera-Verdugo:2020kzc}. 

In this work, we extend the application of LTD to a collection of multiloop topologies that first appear at four loops, and also 
include nonplanar diagrams. All these topologies are unified and their LTD representation 
describes at once the opening of any four-loop scattering amplitude to nondisjoint trees.

\section{Loop-Tree Duality}

A generic $L$-loop scattering amplitude with $N$ external legs, $\{p_j\}_N$,
is encoded in the Feynman representation as an integral in the Minkowski space of the $L$ loop momenta, $\{\ell_s\}_L$, 
over the product of Feynman propagators, $G_F(q_i) = (q_i^2-m_i^2+\ii)^{-1}$, 
and numerators given by the Feynman rules of the specific theory,
\begin{align}
\mathcal{A}_N^{(L)}(1,\ldots, n) = \int_{\ell_1, \ldots, \ell_L} \mathcal{A}_F^{(L)}(1,\ldots, n) \, ,
\end{align}
with
\begin{align}\label{eq:amplitude}
&\mathcal{A}_F^{(L)}(1,\ldots, n)  = \mathcal{N}( \{ \ell_s\}_L,  \{ p_j\}_N) \, G_F(1,\ldots, n) \, .
\end{align}
The integration measure in dimensional regularization~\cite{Bollini:1972ui, tHooft:1972tcz}
reads 
\begin{align}\
\int_{\ell_s} = -\imath \mu^{4-d} \int \frac{d^d\ell_s}{(2\pi)^d}, 
\end{align}
with $d$ the number of space-time dimensions.
In \Eq{eq:amplitude}, we have introduced a shorthand notation to denote the product of Feynman propagators of one set 
that depends on a specific loop momentum or the union of several sets that depend on independent linear combinations 
of the loop momenta, i.e.
\begin{align}
G_F(1,\ldots, n) = \prod_{i\in 1\cup\ldots \cup n} \left( G_F(q_i) \right)^{a_i} \, ,
\end{align}
with $a_i$ arbitrary powers. It is important to remark that from now on the powers $a_i$ will appear only implicitly.
Also, the LTD representations that will be presented do not require to detail the internal configuration of each set.

The LTD representation is obtained by integrating out one degree of freedom per loop through the Cauchy residue theorem.
This results in a modification of the infinitesimal complex prescription of the Feynman propagators~\cite{Catani:2008xa}, that needs to be considered 
carefully to preserve the causal structure of the amplitude. In the context of multiloop scattering amplitudes, 
the LTD representation is written in terms of nested residues~\cite{Verdugo:2020kzh}
\begin{align}
\label{eq:nested}
&\mathcal{A}_D^{(L)}(1,\ldots, r; r+1,\dots, n)  
=-2\pi \imath \sum_{i_r \in r} {\rm Res} (\mathcal{A}_D^{(L)}(1, \ldots, r-1;r, \ldots, n), {\rm Im}(\eta\cdot q_{i_r})<0)\, , 
\end{align}
starting from 
\begin{align}
&\mathcal{A}_D^{(L)}(1; 2, \ldots, n)  
=-2\pi \imath \sum_{i_1 \in 1} {\rm Res} (\mathcal{A}_F^{(L)}(1, \dots, n), {\rm Im}(\eta\cdot q_{i_1})<0)\, , 
\end{align}
where $\mathcal{A}_F^{(L)}(1, \dots, n)$ is the integrand in the Feynman representation, \Eq{eq:amplitude}.
The Cauchy countours are always closed on the lower half plane such that the poles with negative 
imaginary components are selected. This is implemented through the future-like vector $\eta$ that selects which
components of the loop momenta are integrated. 
The usual choice is $\eta^{\mu}=(1,{\bf 0})$, which is equivalent to integrate out the loop energies 
and has some advantages because the remaining integration domain is Euclidean.
The LTD representations presented in the following are, however, independent of the coordinate system. 

The internal structure of $\mathcal{A}_F^{(L)}$ is implicitly specified via the overall tagging of the different
sets of internal momenta. In \Eq{eq:nested}, all sets before the semicolon are linearly independent and each of them contains one
propagator which has been set on shell, while all propagators belonging to sets after the semicolon remain off shell. 
The sum over all possible on-shell configurations in $\mathcal{A}_D^{(L)}$ is understood through the sum of residues.
For example, the LTD representation of the multi-banana or MLT topology has the very compact and symmetric form~\cite{Verdugo:2020kzh}
\begin{align}
\label{eq:mastermlt}
& {\cal A}^{(L)}_{\rm MLT} (1, \ldots, L+1) = \int_{\ell_1, \ldots, \ell_L} \sum_{i=1}^{L+1}\mathcal{A}_D^{(L)}(1,\ldots, i-1, \overline{i+1}, \ldots, \overline{L+1};  i) \, . 
\end{align}
The bars in \Eq{eq:mastermlt} indicate a reversal of momentum flow, $q_{{\overline i}_s} = -q_{i_s}$, which is necessary to
preserve causality.  More details can be found in Ref.~\cite{Verdugo:2020kzh,Aguilera-Verdugo:2020kzc,Aguilera-Verdugo:2020nrp}.

\section{The N$^4$MLT universal topology}

In this work, we study the multiloop topologies that appear for the first time at four loops. 
They are characterized by multiloop diagrams with $L+4$ and $L+5$ sets of propagators.
According to the classification scheme in Ref.~\cite{Verdugo:2020kzh}, 
they correspond to the next-to-next-to-next-to maximal loop topology (N$^3$MLT) 
and next-to-next-to-next-to-next-to maximal loop topology (N$^4$MLT). 
Actually, N$^4$MLT embraces in a natural way all N$^{k-1}$MLT configurations, with $k\le 4$.

\begin{figure}[t!]
\includegraphics[scale=0.8]{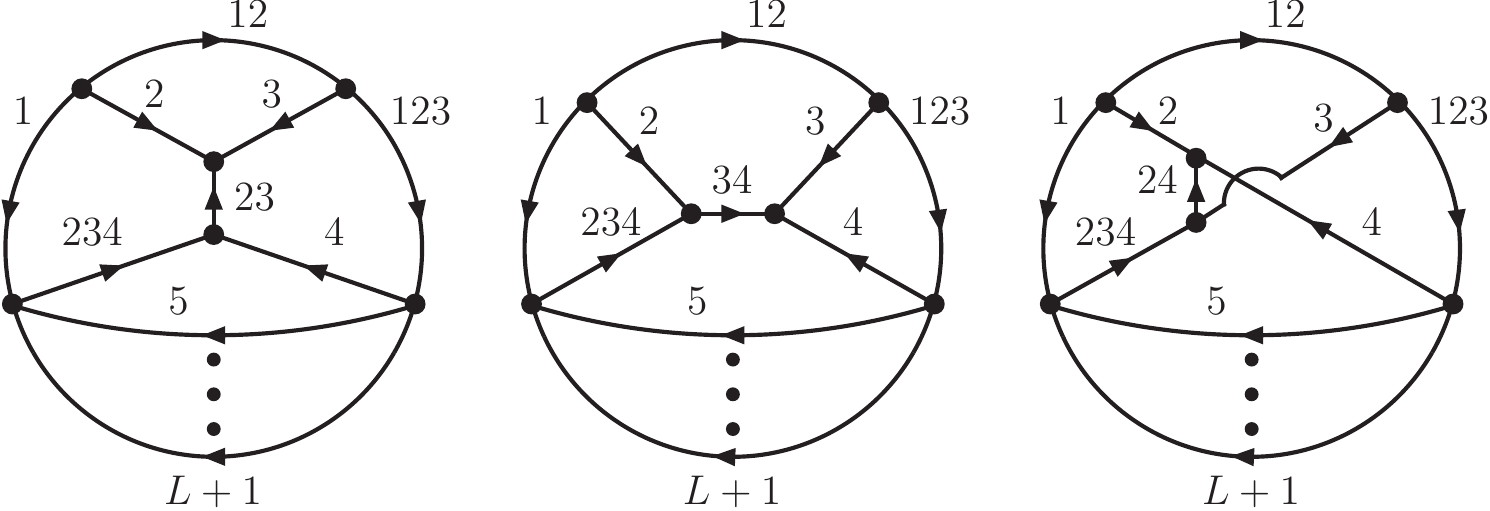}
\centering
\caption{Diagrams of the $\rm N^4MLT$ family. 
The diagram on the l.h.s. corresponds to the $t$ channel, the diagram on the center is the $s$ channel 
and the diagram on the r.h.s. corresponds to the $u$ channel. An arbitrary number of external particles (not shown) are attached. 
\label{stu-diagrams}}
\end{figure}

This arrangement allows to restrict the overall assessment to the N$^4$MLT family that consists of three main topologies. 
These topologies were checked with \verb|QGRAF|~\cite{Nogueira:1991ex} and are shown in Fig.~\ref{stu-diagrams}. 
Two of them are planar and one is nonplanar.
Nicely, we observe the similarity of these topologies with the insertion of a four-point subamplitude
with trivalent vertices into a larger topology.
Therefore, in order to achieve a unified description the three N$^4$MLT topologies are interpreted as the 
$t$-, $s$- and $u$-kinematic channels, respectively, of a {\it universal topology}.

The three topologies contain $L+4$ common sets of propagators, and one extra set which is different for each of them.
Each of the first $L$ sets depends on one characteristic loop momentum $\ell_s$ and the momenta of their propagators have the form
$q_{i_s}=\ell_s + k_{i_s}$. The remaining four common sets are established as linear combinations of the loop momenta, explicitly
\begin{align}
q_{i_{(L+1)}}       & = - \sum_{s=1}^L \ell_s + k_{i_{(L+1)}} \, , \quad q_{i_{12}}    = -  \ell_1 - \ell_2 +k_{i_{12}} \, , \nn   \\
q_{i_{123}} & = - \sum_{s=1}^3 \ell_s +k_{i_{123}} \, , \quad 
q_{i_{234}}    = - \sum_{s=2}^4 \ell_s   +k_{i_{234}} \, ,
\label{eq:momenta1} 
 \end{align}
with $k_{i_s}, k_{i_{(L+1)}}, k_{i_{12}}, k_{i_{123}}$ and $k_{i_{234}}$ linear combinations of external momenta. 
The extra sets are the distinctive key to each of the channels in {\it the universal topology}. 
We identify the momenta of their propagators as different linear combinations of $\ell_2$, $\ell_3$ and $\ell_4$, writing them as
\begin{align}
q_{i_{rs}} & = -  \ell_r - \ell_s +k_{i_{rs}} \, , \quad r, s \in \{2,3,4\}\, .
\label{eq:momenta2} 
\end{align} 

To assemble the three N$^4$MLT channels into a single topology, we define the current $J$ that includes the three different type of sets,
\beq
J \equiv 23 \cup 34 \cup 24~.
\eeq
Notice that due to momentum conservation, the three subsets cannot contribute to the same individual Feynman diagram 
but they all contribute at amplitude level. 
Relying on the development of this framework, the Feynman representation of the N$^4$MLT universal topology can be expressed as
\begin{align}
\mathcal{A}_{\rm N^4MLT}^{(L)} (1, \ldots, L+1, 12, 123, 234, J) = \int_{\ell_1,\dots ,\ell_L}\mathcal{A}_F^{(L)}(1,\dots, L+1,12,123,234,J)~.
\end{align}

\begin{figure}[t!]
\includegraphics[scale=0.8]{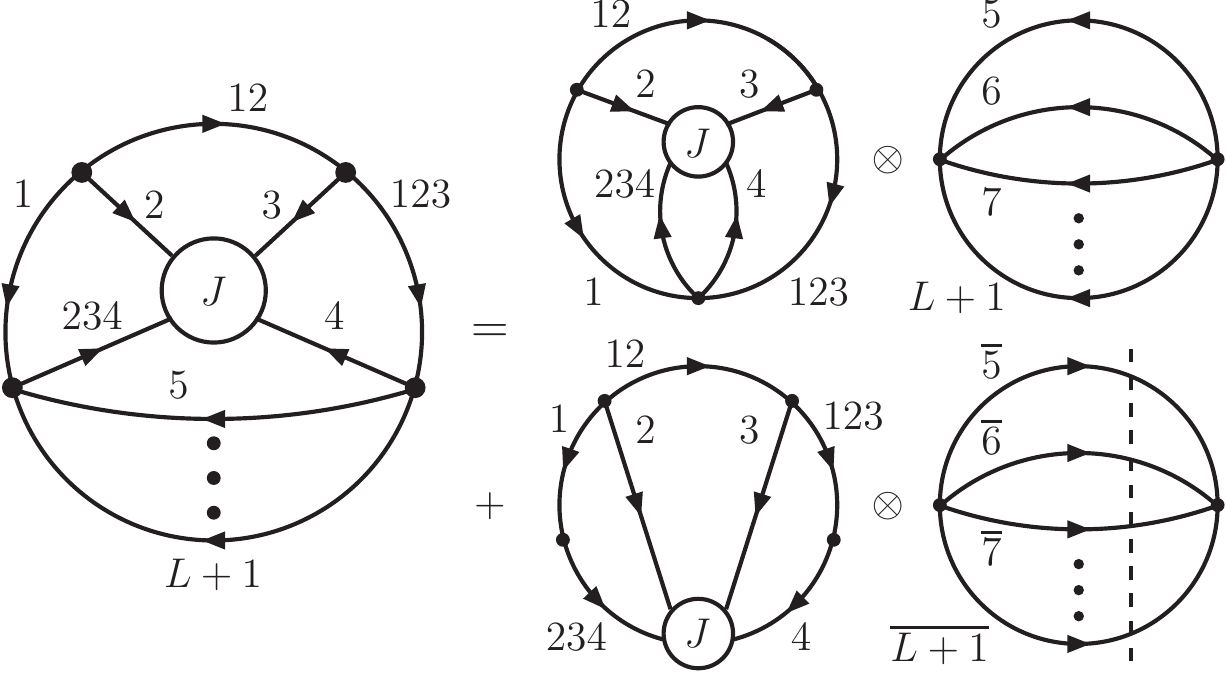}
\centering
\caption{Diagrammatic representation for the factorized opening of the multiloop 
N$^4$MLT {\it universal topology}. Only the on-shell cut of the last MLT-like subtopology 
with reversed momentum flow is shown.
\label{fig:master}}
\end{figure}

The dual opening of this topology fulfills a factorization identity in terms of convoluted subtopologies,
similar to those presented in Ref.~\cite{Verdugo:2020kzh} for NMLT and N$^2$MLT, i.e.
\begin{align}
\label{eq:master}
{\cal A}^{(L)}_{\rm N^4MLT} & (1, \ldots, L+1, 12, 123, 234, J) \nn \\
& =  {\cal A}^{(4)}_{\rm N^4MLT} (1, 2, 3, 4, 12, 123, 234, J) \otimes {\cal A}_{\rm MLT}^{(L-4)} (5, \dots, L+1) \nn \\
& + {\cal A}^{(3)}_{\rm N^2MLT} (1\cup 234, 2, 3, 4\cup 123, 12, J) \otimes {\cal A}_{\rm MLT}^{(L-3)} (\overline 5, \dots, \overline{L+1} )~. 
\end{align} 
The convolution symbol indicates that each of the convoluted components is open independently, whereas
the on-shell conditions from all components act together on the numerator and the propagators that remain off-shell.
An essential constraint to be meet by the selected on-shell propagators concerns the non-feasibility of generating disjoint trees in the dual opening.
In order to make the notation more readable, ${\cal A}^{(L)}_{{\rm N}^{k-1}{\rm MLT}}$ will refer in the following to the integrand 
of the corresponding topology in the LTD representation; integration over the $L$ loop momenta will be implicitly understood. 

The factorization identity in \Eq{eq:master} is the main result of this paper, and it is the universal identity that opens 
any multiloop N$^4$MLT topology to nondisjoint trees. It also enables 
to infer the causal structure of the complete topology by exploring the causal behaviour of its subtopologies.
Let us emphasize that this identity is valid regardless of the internal configuration, i.e. numerators, 
multiple-power propagators and number of external particles, because the residue operator is implicitly considered. 
In addition, since it properly accounts for all N$^{k-1}$MLT configurations with $k\le 4$, 
it is the only master expression required to open to nondisjoint trees any scattering amplitude of up to four loops.
Beyond four loops, new topologies arise which, for consistency, will include this universal topology as a particular case.

A graphical interpretation of the factorization identity is shown in Fig.~\ref{fig:master}. 
The term $\mathcal{A}^{(4)}_{\rm N^4MLT}$ on the r.h.s. of \Eq{eq:master} 
considers all possible configurations with four on-shell 
propagators in the sets $\{1, 2, 3 ,4,12,123,234,J \}$, 
while ${\cal A}^{(3)}_{\rm N^2MLT}$ in the second term
assumes three on-shell conditions under the constraints explained below.
The term ${\cal A}_{\rm MLT}^{(L-4)} (5, \dots, L+1)$ is open according to the MLT opening presented in Ref.~\cite{Verdugo:2020kzh,Aguilera-Verdugo:2020nrp},
and in ${\cal A}_{\rm MLT}^{(L-3)} (\overline 5, \dots, \overline {L+1})$ all the momentum flows are reversed and all the sets contain one on-shell propagator.  
The reversion is imposed by the fact that, in the absence of propagators in the sets $\{12,123, 234,J \}$, 
the master opening in \Eq{eq:master} should coincide with the MLT opening.

 The factorization identity in \Eq{eq:master} has been first tested by explicitly computing the nested residues on the l.h.s.
for specific internal configurations, and then confronting the result with the unfolding (according to the known expressions 
of the corresponding subtopologies) of the Ansatz on the r.h.s., which is motivated by the graphical interpretation.  
In order to prove that \Eq{eq:master} holds for an arbitrary number of loops, we decomposed the N$^4$MLT into two parts: the upper part in the l.h.s. of Fig.~\ref{fig:master} containing the current $J$, which encodes the novel topological complexity arising in this class of diagrams, and the lower one, which represents a known MLT-like component. The LTD representation of the topological complexity part is computed at four loops for each configuration described by the current $J$, or equivalently, by considering the MLT-like sector as a single internal line. Then, we just need to rely on the MLT formula presented in \Eq{eq:mastermlt} to complete the calculation and obtain the all-order expression.

We have to mention certain arbitrariness in the expression~(\ref{eq:master}) due to the freedom in the order of the nested application of the Cauchy residue theorem. Although there are at least $L!$ possibilities, all potential LTD representations are, however, equivalent and lead to the same causal expression in terms of dual propagators~\cite{Aguilera-Verdugo:2020kzc} (see Sec.\ref{sec:causal}).

\begin{figure}[t!]
\includegraphics[scale=0.8]{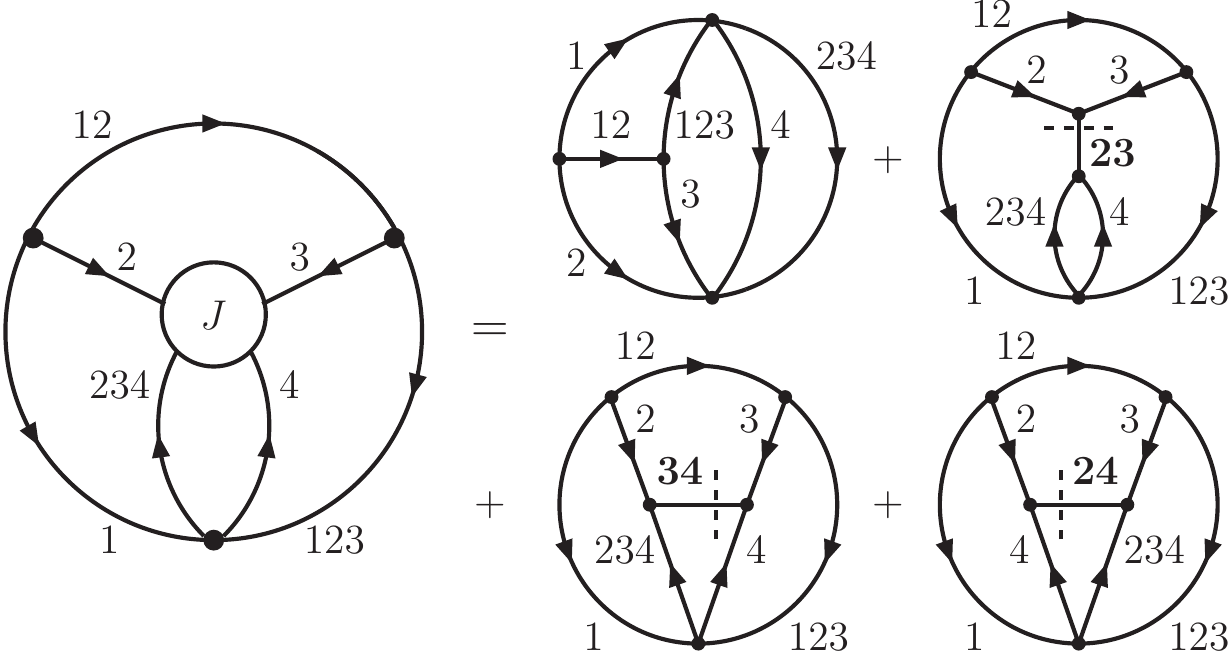}
\centering
\caption{Diagrammatic representation of the four-loop subtopology ${\cal A}^{(4)}_{\rm N^4MLT} (1, 2, 3, 4, 12, 123, 234, J)$. 
The dashed lines and bold labels indicate that propagators in the $J$-sets are set on shell in either of the two momentum flows.
\label{fig:ec1}}
\end{figure}

The four-loop subtopology in \Eq{eq:master} is opened as well through a factorization identity which is written in terms of known subtopologies, 
\begin{align}
\label{eq:fourloop}
\mathcal{A}_{\rm N^4MLT}^{(4)} (1, 2, 3, 4, 12, 123, 234, J) 
& = \mathcal{A}_{\rm{N^2MLT}}^{(4)}(1, 2, 3, 4, 12, 123, 234) \otimes \mathcal{A}^{(0)}(J)  \nn \\ 
& +\sum_{{\bf s}\in J}  \mathcal{A}_D^{(4)}(1, 2, 3, 4, 12, 123, 234, {\bf s}) \, ,
\end{align}
The diagrammatic representation of \Eq{eq:fourloop} is depicted in Fig.~\ref{fig:ec1}. 
The first term on the r.h.s. of \Eq{eq:fourloop} consists of a four-loop $\rm{N^2MLT}$ subtopology and 
describes dual trees where all the propagators with momenta in $J$ remain off shell, corresponding to the first diagram on the r.h.s. of Fig.~\ref{fig:ec1}.
The second term on the r.h.s. of \Eq{eq:fourloop} collects contributions where propagators in either $23$, $34$ or $24$ are set on shell. 
These dual trees are therefore specific to the $t$, $s$ and $u$ channels, whose explicit expressions are presented below.
We use the bold symbol ${\bf s}$ to clearly indicate that these contributions are those containing on-shell propagators in the $J$-sets.

\begin{figure}[t!]
\includegraphics[scale=0.8]{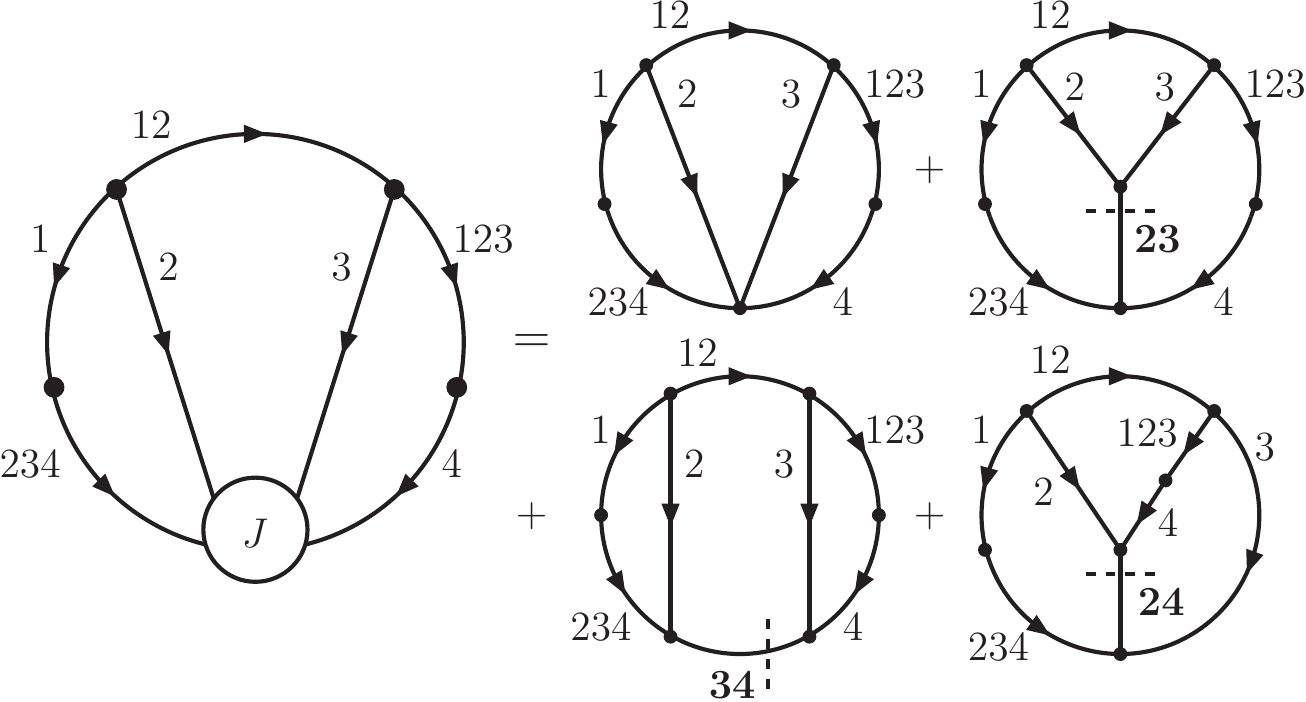}
\centering
\caption{Diagrammatic representation of the three-loop subtopology ${\cal A}^{(3)}_{\rm N^2MLT} (1\cup 234, 2, 3, 4\cup 123, 12, J)$.
The dashed lines and bold labels indicate that propagators in the $J$-sets are set on shell in either of the two momentum flows. 
\label{fig:ec2}}
\end{figure}

The three-loop subtopology on the r.h.s. of \Eq{eq:master} is also opened in terms of known subtopologies through the factorized identity
\begin{align}
\label{eq:threeloop}
{\cal A}^{(3)}_{\rm N^2MLT} (1\cup 234, 2, 3, 4\cup 123, 12, J) 
& = \mathcal{A}_{\rm{NMLT}}^{(3)}(1\cup234, 2, 3, 4\cup 123, 12)\otimes \mathcal{A}^{(0)}(J) \nn \\ 
& + \sum_{{\bf s}\in J}  \mathcal{A}_D^{(3)}(1, 2, 3, 4, 12, 123, 234, {\bf s})  \, ,
\end{align} 
which has a similar structure to \Eq{eq:fourloop}.
The diagrammatic representation of \Eq{eq:threeloop} is depicted in Fig.~\ref{fig:ec2}.
Similarly to Fig.~\ref{fig:ec1}, the first diagram on the r.h.s. of Fig.~\ref{fig:ec2}, which represents 
the first term on the r.h.s. of \Eq{eq:threeloop}, is a three-loop NMLT subtopology and all the propagators in $J$
are off shell, while the remaining three diagrams are specific to each of the three channels.
The NMLT subtopology is made up of 7 subsets of momenta grouped into 5 sets as follows $\left\lbrace 1\cup234, 2, 3, 4\cup 123, 12 \right\rbrace$.
This construction prevents, for example, that propagators in the sets $1$ and $234$ are set on shell simultaneously.


Turning back into Eqs.~(\ref{eq:fourloop}) and (\ref{eq:threeloop}) in a more detailed way, the first terms on the r.h.s. of both equations 
are composed of dual contributions where all the propagators in $J$ remain off shell. These $J$-propagators act as spectators
in relation to the opening of the accompanying subtopology, and can eventually be replaced by a contact interaction to deduce the opening
rule of these contributions. 

Specifically, the four-loop N$^2$MLT subamplitude in \Eq{eq:fourloop} is represented by 
\begin{align} \label{eq:NNMLT}
\mathcal{A}_{\rm{N^2MLT}}^{(4)}(1, 2, 3, 4, 12, & 123, 234)  
=\mathcal{A}_{\rm{NMLT}}^{(3)}(1, 2, 3, 12, 123, 234)\otimes \mathcal{A}_{\rm{MLT}}^{(1)}(4, 234)\nn \\
&+\left[ \mathcal{A}_{\rm{MLT}}^{(2)}(1, 2, 12)+\mathcal{A}^{(2)}_{\rm MLT}(1, \overline{3}) \right]\otimes \mathcal{A}^{(2)}_{\rm MLT}(\overline{4},\overline{234})\nn \\ & +\left[ \mathcal{A}_{\rm{MLT}}^{(2)}(\overline{123}, \overline{3}, 12)+\mathcal{A}^{(2)}_{\rm MLT}(2, \overline{123}) \right]\otimes \mathcal{A}^{(2)}_{\rm MLT}(4, 234) \, . 
\end{align}
All the MLT subamplitudes that involve a number of loops equal to the number of sets require to set on shell propagators in all the sets. 
The rest of NMLT and MLT subtopologies are open according to known expressions \cite {Verdugo:2020kzh}. 
To simplify the presentation, we have omitted in \Eq{eq:NNMLT} the explicit reference to the sets with all their propagators off shell; 
for instance, the element $\mathcal{A}^{(2)}_{\rm MLT}(1,\overline{3})\otimes\mathcal{A}^{(2)}_{\rm MLT}(\overline{4},\overline{234})$ 
must be interpreted as
\begin{align}
\mathcal{A}^{(2)}_{\rm MLT}& (1,\overline{3})\otimes\mathcal{A}^{(2)}_{\rm MLT}(\overline{4},\overline{234}) = \mathcal{A}_D^{(4)}(1,\overline{3},\overline{4},\overline{234}; 2,12,123) \, .
\end{align} 
This notation will be used in the following; the omitted sets are understood to be off shell. 

The three-loop ${\rm NMLT}$ subtopology in \Eq{eq:threeloop} is generated from 7 subsets clustered 
as $\{ 1\cup 234, 2, 3, 4\cup 123, 12 \}$ and its LTD representation is
\begin{align}
  \mathcal{A}_{\rm{NMLT}}^{(3)}(1\cup 234, 2, 3, 4\cup 123, 12) 
& =\mathcal{A}_{\rm{MLT}}^{(2)}(1\cup 234, 2, 12) \otimes \mathcal{A}_{\rm{MLT}}^{(1)}(3, 4\cup 123) \nn \\
& +\mathcal{A}_{\rm{MLT}}^{(1)}(1\cup 234, 2) \otimes \mathcal{A}^{(2)}_{\rm MLT}(\overline{3}, \overline{4}\cup \overline{123}),
\end{align}
where the first term is a convolution of two MLT subtopologies, and in the second term all the propagators in the set $12$ are off shell.

\subsection{The t channel}

The second terms in Eqs.~(\ref{eq:fourloop}) and (\ref{eq:threeloop}) distinguish the dual configurations arising 
for each of the three channels when propagators with momenta in $J$ are set on shell. 
We begin analyzing the dual terms related exclusively to the topology known as $t$ channel from Fig.~\ref{stu-diagrams} (left).
There is one four-loop subtopology
\begin{align}
\mathcal{A}_D^{(4)}(1, 2, 3, 4,  12, &123, 234, {\bf 23})   
=\left[ \left( \mathcal{A}_{\rm{MLT}}^{(2)}(\overline{123}, \overline{3}, 12) + \mathcal{A}^{(2)}_{\rm MLT}(2, \overline{123}) \right) \otimes \mathcal{A}^{(1)}_D(23)  \right.  \nn \\
&+\left. \left( \mathcal{A}_{\rm{MLT}}^{(2)}(1, 2, 12) + \mathcal{A}^{(2)}_{\rm MLT}(1, \overline{3})\right)\otimes \mathcal{A}^{(1)}_D(\overline{23}) \right] \otimes \mathcal{A}^{(1)}_{\rm MLT}(4, 234)~, 
\label{eq:tone}
\end{align}
and one three-loop subtopology
\begin{align}
\mathcal{A}_D^{(3)}(1, 2, 3, 4, 12, &123, 234, {\bf 23}) 
=\left[ \mathcal{A}_{\rm{MLT}}^{(2)}(\overline{4}\cup \overline{123}, \overline{3}, 12) + \mathcal{A}^{(2)}_{\rm MLT}(2, \overline{4}\cup \overline{123})  \right] \otimes \mathcal{A}^{(1)}_D(23) \nn \\
&+ \left[ \mathcal{A}_{\rm{MLT}}^{(2)}(1\cup 234, 2, 12) + \mathcal{A}^{(2)}_{\rm MLT}(1\cup 234, \overline{3})  \right] 
\otimes \mathcal{A}^{(1)}_D(\overline{23})~, 
\label{eq:ttwo}
\end{align}
that contribute to Eqs.~(\ref{eq:fourloop}) and (\ref{eq:threeloop}), respectively.
Notice that both expressions, \Eq{eq:tone} and \Eq{eq:ttwo},
contain on-shell propagators in the set $23$, and we have contributions
with the original momentum flow, $23$, and the reversed one $\overline{23}$.

\subsection{The s channel}

In order to obtain the terms that characterize the $s$ channel shown in Fig.~\ref{stu-diagrams} (center), 
we set on shell a propagator in the set $34$. The four-loop subtopology is given by \\ \\
\begin{align}
&\mathcal{A}_D^{(4)}(1, 2, 3, 4, 12, 123, 234, {\bf 34})  \nn \\
&=\left[ \mathcal{A}_{\rm{NMLT}}^{(3)}(1, 2, 3, 12, 123) +\mathcal{A}^{(3)}_{\rm MLT}(\overline{3}, \overline{123},234) \right. \nn \\
&+ \mathcal{A}_{\rm{MLT}}^{(2)}(1, 2, 12)\otimes \mathcal{A}^{(1)}_D(\overline{4}) + \left. \mathcal{A}_{\rm{MLT}}^{(1)}(3\cup \overline{4}, 123)\otimes \mathcal{A}^{(2)}_{\rm MLT}(\overline{12}, 234)  \right] \otimes \mathcal{A}^{(1)}_D(\overline{34}) \nn \\
&+\left[\mathcal{A}_{\rm{MLT}}^{(1)}(3\cup \overline{4}, 123)\otimes  \mathcal{A}^{(2)}_{\rm MLT}(1, \overline{234})  \right. + \left.\mathcal{A}_{\rm{MLT}}^{(1)}(1, 2\cup \overline{234})\otimes \mathcal{A}^{(2)}_{\rm MLT}(4, \overline{123}) \right] \otimes \mathcal{A}^{(1)}_D(34) \, . 
\end{align}
This expression is more involved than the corresponding one in the $t$ channel, because 
the loop momentum $\ell_4$ is now present in three sets, while in the $t$ channel $\ell_4$ is found in two sets only. 

By contrast, for the three-loop subamplitude we observe  a very symmetric structure which allows to avoid any momentum flow reversion. In this case, we end up with an expression that only depends on the original momentum flow of the set $34$,
\begin{align}
&\mathcal{A}_D^{(3)}(1, 2, 3, 4, 12, 123, 234, {\bf 34}) 
=\mathcal{A}_{\rm{MLT}}^{(1)}(1\cup 234, 2) \otimes \mathcal{A}_{\rm{MLT}}^{(1)}(3, 4\cup 123) \otimes \mathcal{A}^{(1)}_D(34)\, .
\end{align}
Given the structure of this subtopology, it is manifest that propagators in 12 and 34 cannot become on shell simultaneously 
without generating a disjoint tree, as expected from Fig.~\ref{fig:ec2}. \\

\subsection{The u channel}

Moving on to the last terms associated to the nonplanar topology known as $u$ channel (Fig.~\ref{stu-diagrams} (right)), 
the LTD representation of the four-loop subamplitude with on-shell propagators in the set $24$ reads
\begin{align}
\mathcal{A}_D^{(4)}&(1, 2, 3, 4, 12, 123, 234, {\bf 24}) \nn \\ 
&=\left[ \mathcal{A}_{\rm{NMLT}}^{(3)}(1, 2, 3, 12, 123) \right. + \mathcal{A}_{\rm{MLT}}^{(1)}(1, 2\cup \overline{4})\otimes \mathcal{A}^{(2)}_{\rm MLT}(\overline{123},234) \nn \\
&+ \left. \mathcal{A}_{\rm{MLT}}^{(1)}(3\cup \overline{234}, 123)\otimes \mathcal{A}^{(2)}_{\rm MLT}(1, \overline{4}) \right] \otimes \mathcal{A}^{(1)}_D(\overline{24}) \nn \\
&+\left[ \mathcal{A}_{\rm{MLT}}^{(2)}(1,2,12)\otimes \mathcal{A}^{(1)}_{\rm MLT}(\overline{234})+ \mathcal{A}^{(3)}_{\rm MLT}(4, \overline{3},\overline{123}) \right. \nn \\
&+ \left.   \mathcal{A}_{\rm{MLT}}^{(1)}(3\cup \overline{234}, 123)\otimes  \mathcal{A}^{(2)}_{\rm MLT}(4,\overline{12}) \right] \otimes \mathcal{A}^{(1)}_D(24) \, .
 \end{align}
This subtopology is also not as compact as the expression for the $t$ channel because $\ell_4$ is also present in three different sets. 
For the three-loop subamplitude, we find, 
\begin{align}
\mathcal{A}_D^{(3)}(1, 2, 3, 4, 12, 123, 234, {\bf 24}) &= \left[ \mathcal{A}^{(2)}_{\rm MLT}(1\cup 234, \overline{4}) +\mathcal{A}^{(2)}_{\rm MLT}(234, \overline{123}) \right] \otimes \mathcal{A}^{(1)}_D(\overline{24})
\nn  \\
&+\left[ \mathcal{A}_{\rm{MLT}}^{(2)}(1\cup 234\cup \overline{3}, 2, 12) + \mathcal{A}^{(2)}_{\rm MLT}(\overline{3}, \overline{4} \cup \overline{123}) \right. \nn \\
&+\left. \mathcal{A}^{(2)}_{\rm MLT}(4\cup 123, \overline{12}) + \mathcal{A}^{(2)}_{\rm MLT}(\overline{1}, 123) \right] \otimes \mathcal{A}^{(1)}_D(24)~.
\end{align}

All these results are consistent with the absence of disjoint trees. We would like to comment on the fact that
repeated propagators from selfenergy insertions are treated as single propagators raised to specific powers
and are not considered to generate disjoint trees when the repeated propagator is set on shell.

Notice that the number of trees in the LTD forest can also be computed 
through the combinatorial exercise of selecting, from the full list of sets, all possible subsets of $L$
elements that cannot generate disjoint trees.
For the individual $t$, $s$ and $u$ channels the number of terms calculated in this way are 
$5(8L-17)$, $15(3L-7)$ and $9(5L-11)$, respectively, and $82L-187$ for the 
N$^4$MLT universal topology, in agreement with the number of dual contributions generated by \Eq{eq:master}.
The momentum flows of the on-shell propagators, however, can only be determined through the nested residues.

\section{Causal representations}
\label{sec:causal}

The LTD representation generates in general two kind of singularities, causal thresholds 
that can be interpreted in terms of causality, and noncausal thresholds that are unphysical 
and cancel among different dual terms~\cite{Buchta:2014dfa,Aguilera-Verdugo:2019kbz}.
In Ref.~\cite{Verdugo:2020kzh}, we conjectured that, in fact, the LTD representation
can be rearranged in such a way that noncausal singularities are manifestly absent
in the analytic expression. In this section, we confirm the causal conjecture for the 
N$^4$MLT family and present explicit causal representations of selected configurations 
that can be described through very compact analytic expressions. 

We consider, in the first place, the multi-loop N$^3$MLT configuration with one internal propagator in each loop set and five external momenta
\beq
{\cal A}_{{\rm N}^3{\rm MLT}}^{(L)} (1,\ldots, L+4) = \int_{\ell_1, \cdots, \ell_L}  \mathcal{N}( \{ \ell_s\}_L,  \{ p_j\}_N) \, G_F(1, \ldots, L+1, 12, 123, 234)~,
\label{eq:N3MLTFeynman}
\eeq
where the internal momenta are defined in \Eq{eq:momenta1}. Its LTD representation is obtained through the universal
N$^4$MLT expression in \Eq{eq:master} by considering $J$ as an empty set, or equivalently, by substituting $J$
by a contact interaction. After computing the nested residues and adding them all together, the dual representation reads
\beq
{\cal A}_{{\rm N}^3{\rm MLT}}^{(L)} (1,\ldots, L+4) = 
\int_{\vec \ell_1, \cdots, \vec \ell_L} \frac{{\cal N}_{{\rm N}^3{\rm MLT}} (\{\qon{s}, k_{j,0}\})}{x_{L+4} \left( \prod_{i=1}^{13} \lambda_i^+ \lambda_i^- \right)}~,
\label{eq:N3MLTiscausal}
\eeq
where the integrand is a function of the on-shell energies, $\qon{s} = \sqrt{\qb_s^2 + m_s^2-\ii}$, 
with $s\in \{1, \ldots, L+4\}$, and the energy components of the linear combinations of external momenta, $k_{j,0}$.
This is an integral in the loop three-momenta with integration measure
\beq
\int_{\vec \ell_s} = - \mu^{4-d} \int \frac{d^{d-1}\ell_s}{(2\pi)^{d-1}}~.
\eeq

\begin{figure}[t!]
\includegraphics[scale=0.8]{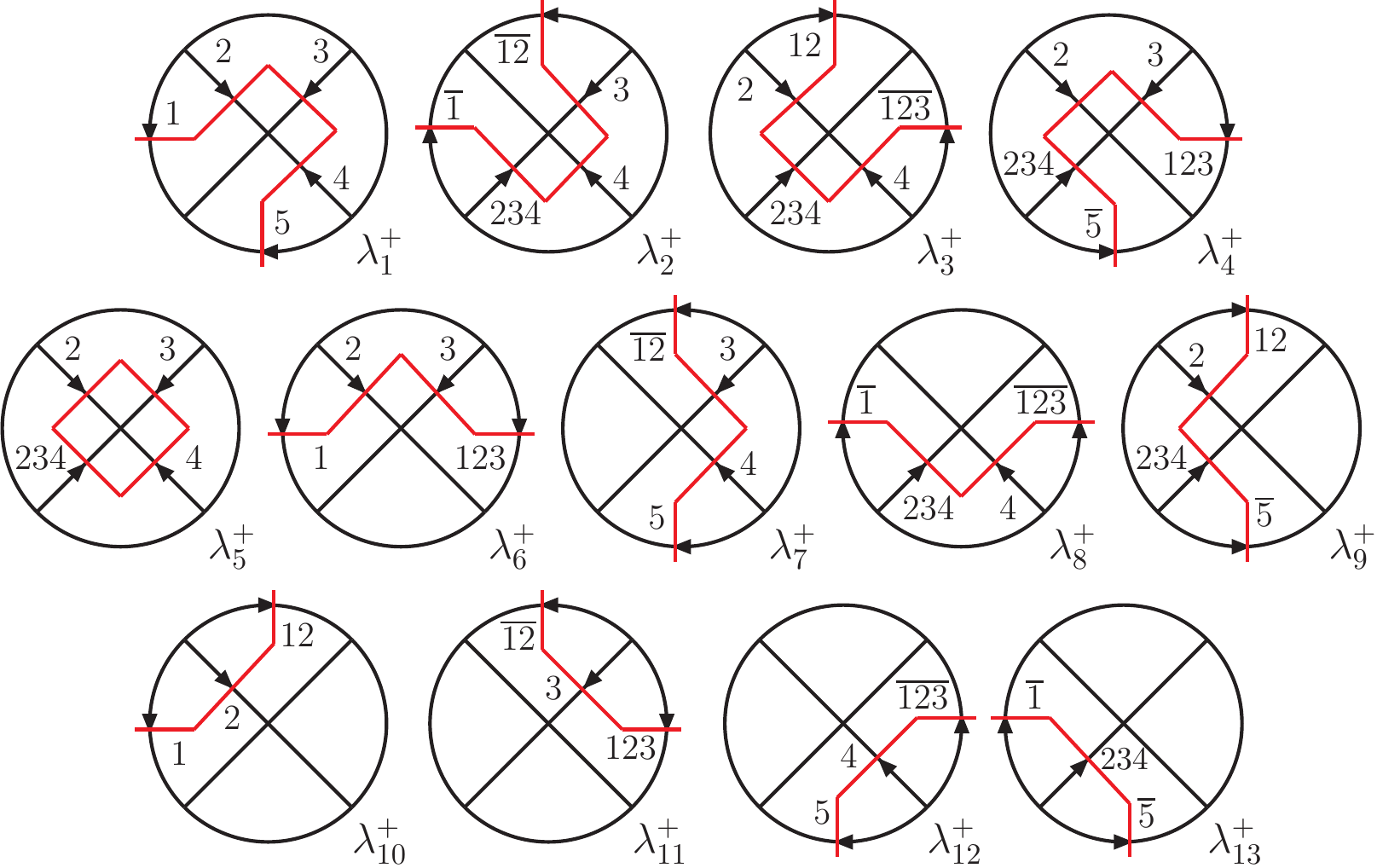}
\centering
\caption{Causal configurations of the ${\rm N^3MLT}$ topology. The $\lambda_i^-$ configurations are obtained 
by reversing the momentum flows of the corresponding $\lambda_i^+$ configurations shown here. The set number $5$ accounts 
for all the propagators in the sets $5$ to $L+1$.
\label{fig:N3MLTcausal}}
\end{figure}

The integrand in \Eq{eq:N3MLTiscausal} is written in terms of 
\beq
x_{L+4} = \prod_{s=1}^{L+4} 2 \qon{s}~,
\eeq
and the thirteen causal denominators
\begin{align}
&   \lambda_1^\pm = \qon{(1,\ldots, L+1)} \pm k_{L+1,0}~, 
&& \lambda_2^\pm = \qon{(1, 12, 3, 4, 234)} \pm (k_{234}-k_{12})_0~, \nn \\
&   \lambda_3^\pm = \qon{(2, 4, 12, 123, 234)} \pm (k_{12}+k_{234}-k_{123})_0~, 
&&   \lambda_4^\pm = \qon{(2, 3, 123, 234, 5, \ldots, L+1)} \pm (k_{123}+k_{234}-k_{L+1})_0~, \nn \\
&  \lambda_5^\pm = \qon{(2,3,4, 234)}  \pm k_{234,0}~, \nn \\
&   \lambda_6^\pm = \qon{(1, 2, 3, 123)} \pm k_{123,0}~, 
&& \lambda_7^\pm = \qon{(12,3,4, \ldots, L+1)} \pm (k_{L+1}-k_{12})_0~, \nn \\
&   \lambda_8^\pm = \qon{(1, 4, 123, 234)} \pm (k_{234}-k_{123})_0~,
&& \lambda_9^\pm = \qon{(2, 12, 234, 5, \ldots, L+1)} \pm (k_{12}+k_{234}-k_{L+1})_0~, \nn \\
&   \lambda_{10}^\pm = \qon{(1, 2, 12)} \pm k_{12,0}~, 
&& \lambda_{11}^\pm = \qon{(12, 3, 123)}  \pm (k_{123}-k_{12})_0~, \nn \\ 
&   \lambda_{12}^\pm = \qon{(123,4, \ldots, L+1)} \pm (k_{L+1} - k_{123})_0~, 
&& \lambda_{13}^\pm = \qon{(1,234,5,\ldots, L+1)} \pm (k_{234}-k_{L+1})_0~, 
\label{eq:lambdaN3MLT}
\end{align}
where we have defined $\qon{(\alpha)} = \sum_{i \in \alpha} \qon{i}$. As explained in Refs.~\cite{Verdugo:2020kzh,Aguilera-Verdugo:2020kzc},
these denominators are causal because they are constructed from sums of on-shell energies exclusively, and they represent 
potential singular configurations in which the momentum flows of the internal propagators are aligned in the same direction.  
The causal denominators appear in pairs because there are two opposite directions to consider for each aligned configuration.
A graphical representation of these causal configurations is shown in Fig.~\ref{fig:N3MLTcausal}.
All other linear combinations of on-shell energies do not have a physical interpretation in terms of causality, 
and are cancelled analytically in the sum over nested residues. 
As a result, the straightforward application of LTD leads directly to an 
expression, \Eq{eq:N3MLTiscausal}, which is manifestly causal. In addition, the causal expression is independent of the 
initial flow assigments of the internal momenta, and of the order of the nested application of the Cauchy residue theorem.

However, the numerator of the integrand, ${\cal N}_{{\rm N}^3{\rm MLT}} (\{\qon{s}, k_{j,0}\})$, 
is a lengthy polynomial in the on-shell and external energies. 
For example, it is a polynomial of degree nine for the scalar integral.
A more suitable causal expression can be obtained 
by reinterpreting  \Eq{eq:N3MLTiscausal} in terms of a number of entangled thresholds 
equal to the difference between the number of propagators and the number of loops by, e.g., analytical 
reconstruction from numerical evaluation over finite fields~\cite{vonManteuffel:2014ixa,Peraro:2016wsq,Peraro:2019svx}
as defined in Ref.~\cite{Aguilera-Verdugo:2020kzc}. The N$^3$MLT expression in \Eq{eq:N3MLTiscausal} 
is analytically reconstructed by matching all combinations of four thresholds that are causally compatible to each other
\beq
{\cal A}_{{\rm N}^3{\rm MLT}}^{(L)} (1,\ldots, L+4) = 
\int_{\vec \ell_1, \cdots, \vec \ell_L}  \frac{1}{x_{L+4} } \sum_\sigma
\frac{{\cal N}_{\sigma(i_1, \ldots ,i_4)} 
(\{\qon{s}, k_{j,0}\})}{\lambda_{\sigma(i_1)} \lambda_{\sigma(i_2)} \lambda_{\sigma(i_3)} \lambda_{\sigma(i_4)}}~,
\label{eq:N3MLTiscausalyes}
\eeq
where $\lambda_{\sigma(i)} \in \{ \lambda_i^\pm\}_{i=1, \ldots,13}$, and the numerators ${\cal N}_{\sigma(i_1, \ldots, i_4)}$ are of the same 
polynomial order as the original numerator in the Feynman representation, e.g., they are constants for scalar integrals. To simplify the discussion, 
we will present explicit results only for scalar integrals because they fully encode all the compatible causal matchings. For example, given a quadratic 
numerator, we can use the identity
\beq
\left( q_{i,0}\right)^2 G_F(q_i) = 1 + \left( \qon{i}\right)^2 G_F(q_i)~.
\label{eq:quadratic}
\eeq
The first term on the r.h.s. generates a scalar integral with one propagator less than the original integral, while the second term introduces 
a factor $\left( \qon{i}\right)^2$ which is not modified by the application of the Cauchy residue theorem. Both integrals, however, are described 
by the same set of causal thresholds. In general, tensor reduction commutes with LTD and can be used to deal with tensor integrals.   

The explicit expression that we obtain for the scalar N$^3$MLT is very compact:
\bea
&& {\cal A}_{{\rm N}^3{\rm MLT}}^{(L)} (1,\ldots, L+4) = \int_{\vec \ell_1, \cdots, \vec \ell_L} \frac{1}{x_{L+4}} 
\bigg[ {\cal F}^{(L+4)}_{(1,5,6,7,10,11,12,13)}  + {\cal F}^{(L+4)}_{(2,5,7,8,11,12,13,10)}  \nn \\
&& \qquad + {\cal F}^{(L+4)}_{(3,5,8,9,12,13,10,11)} + {\cal F}^{(L+4)}_{(4,5,9,6,13,10,11,12)}  
+ L^+_{(6,10,11)} L^-_{(8,12,13)} \nn \\
&& \qquad  + L^-_{(7,11,12)} L^+_{(9,13,10)} + L^+_{10} L^-_{11} L^+_{12} L^-_{13}  + (\lambda_i^+ \leftrightarrow \lambda_i^-)\bigg]~,
\label{eq:N3MLTcausal}
\eea 
where  
\bea
&& {\cal F}^{(L+4)}_{(1,5,6,7,10,11,12,13)} = L^+_1 \left( L^+_5 +L^-_{13} \right) 
\left( L^+_{(6,10,11)} + L^+_{(7,11,12)} + L^+_{10} L^+_{12} \right)~, \nn \\
\eea
with 
\beq
L_i^\pm = \frac{1}{\lambda_i^\pm}~,  \qquad
L_{(i, j, k)}^\pm = L^\pm_i \left(L^\pm_{j} + L^\pm_{k} \right)~. 
\eeq 
The function ${\cal F}^{(L+4)}$ encodes four causal configurations that are obtained by permutation of the arguments.
The number of terms generated by \Eq{eq:master} scales as $3(8L-17)$, i.e. 45 terms at four loops, 
while the number of terms generated by \Eq{eq:N3MLTcausal} equals $98$ regardless of the number of loops.
The numerical performance of \Eq{eq:N3MLTcausal} is, in addition, a factor $2$ to $8$ faster than \Eq{eq:master} at four loops, 
and even three orders of magnitude faster than \Eq{eq:N3MLTiscausal}. 
We have estimated the numerical performance by comparing the timings of evaluating the integrands at $1000$ random points. 
Similar relative timing are observed for the rest of the configurations presented in this paper. 

\begin{figure}[t!]
\includegraphics[scale=0.7]{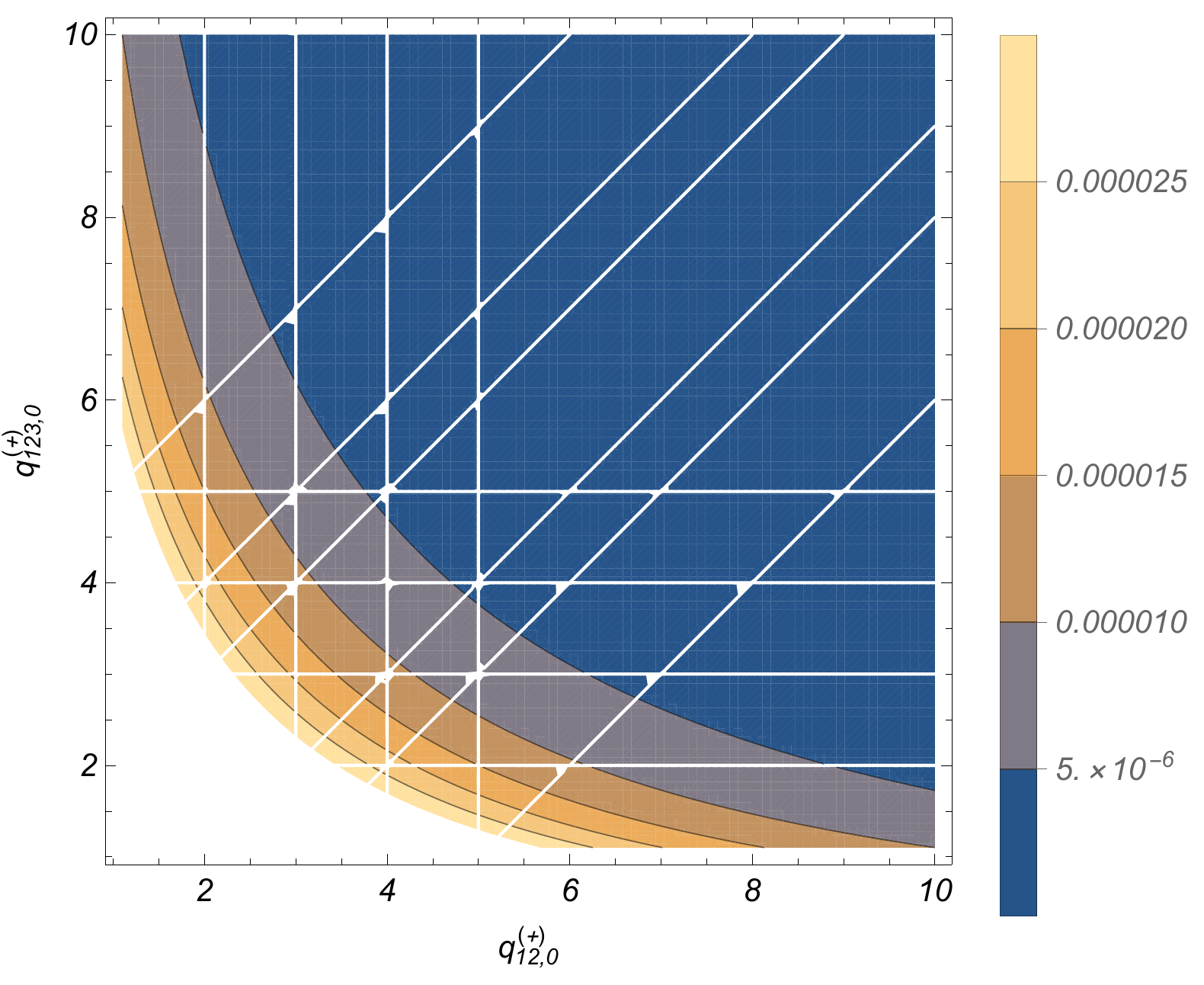}
\centering
\caption{Integrand-level behaviour of the noncausal LTD representation of a four-loop N$^3$MLT diagram. White lines indicate the position of noncausal thresholds.
\label{fig:N3MLTtIntegrand}}
\end{figure}

\begin{figure}[t!]
\includegraphics[scale=0.58]{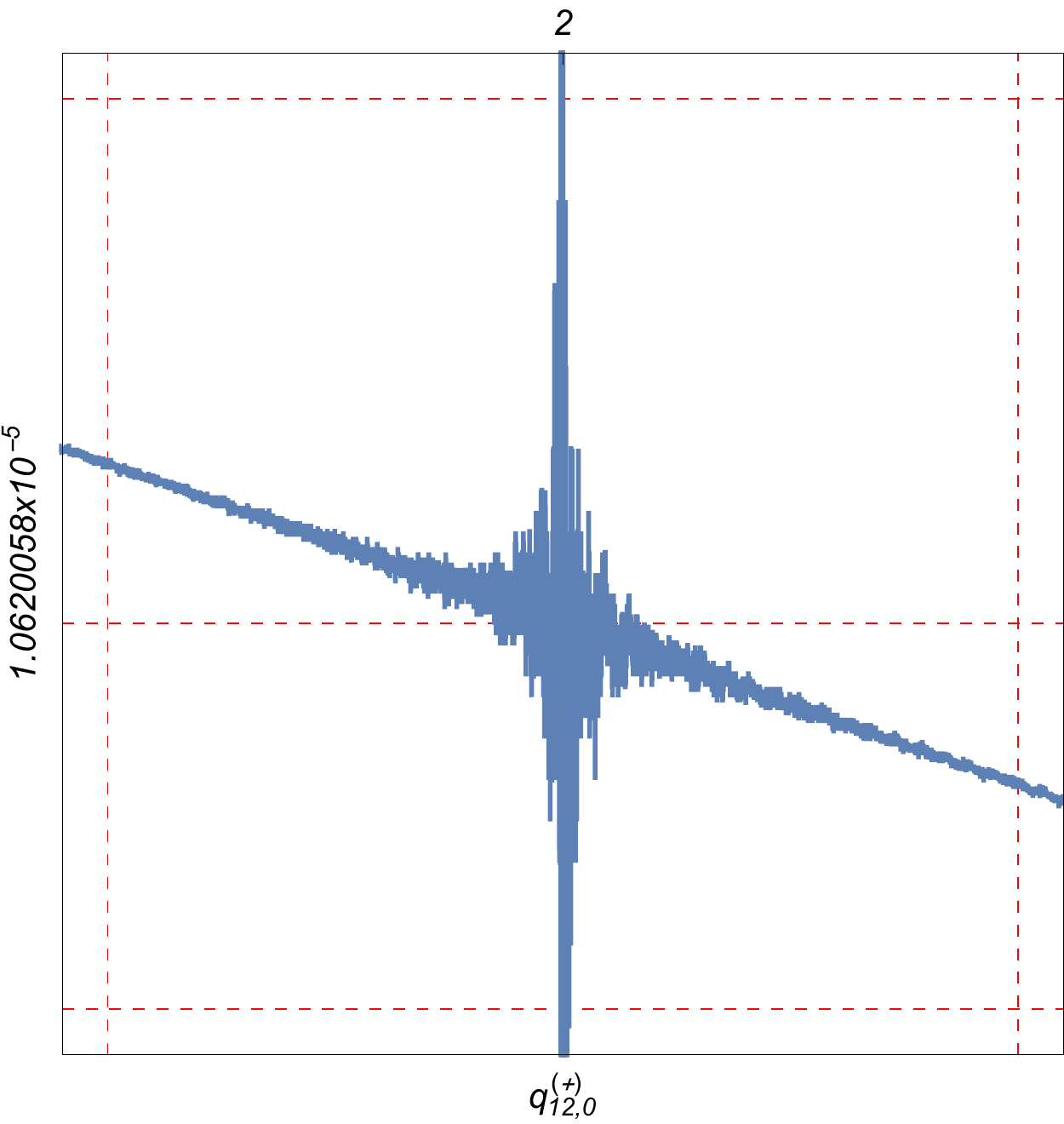} \ \ \includegraphics[scale=0.58]{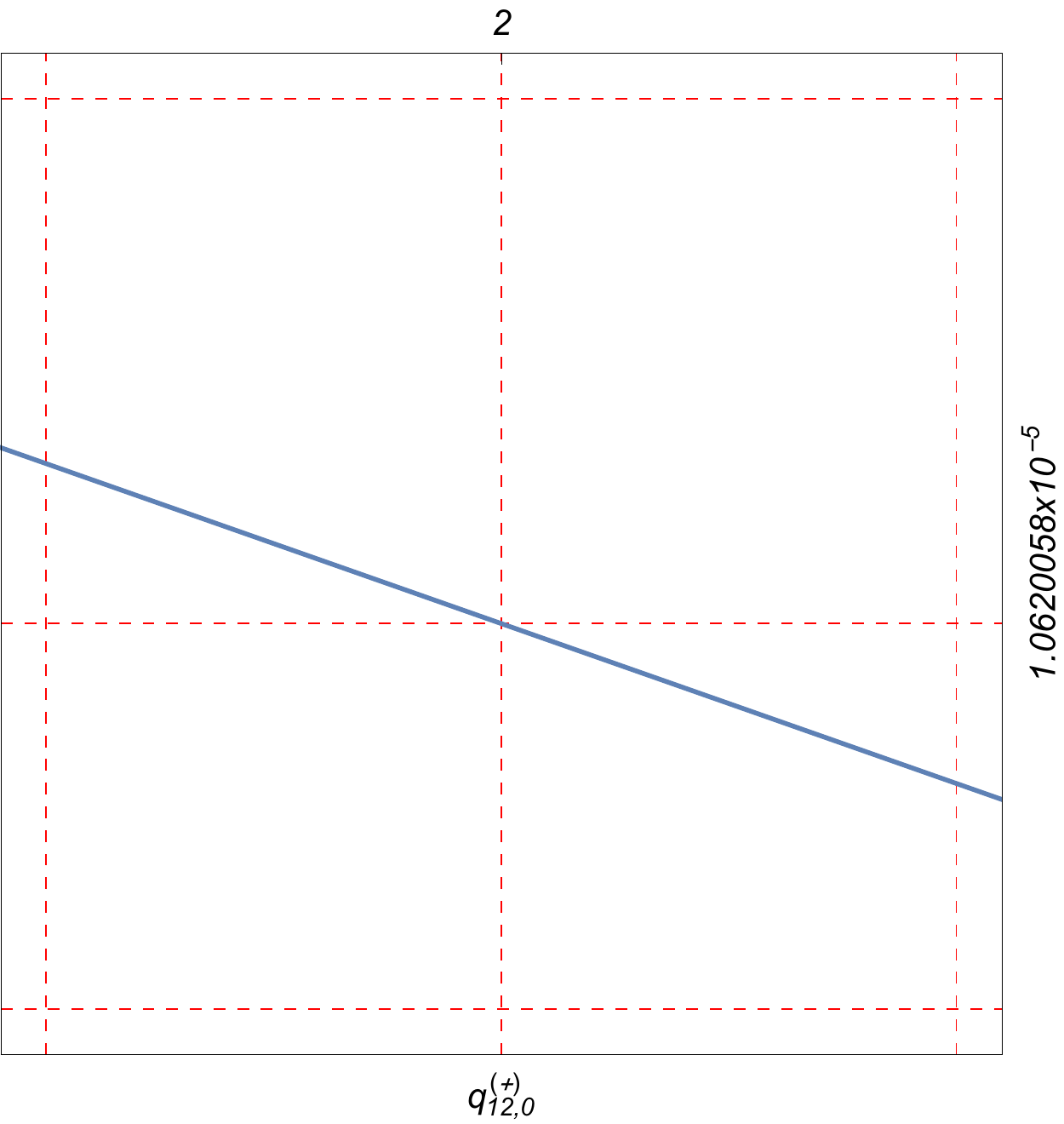}
\centering
\caption{Numerical instabilities of the four-loop N$^3$MLT integrand which arise due to noncausal singularities (left), which are absent in the causal representations (right).
\label{fig:N3MLTtestabilidad}}
\end{figure}

Let us emphasize that the most significant advantage of \Eq{eq:N3MLTcausal} with respect to \Eq{eq:master} stems from the core difference between them, the presence or absence of noncausal singularities. The straightforward application of the nested residue generates multiple threshold singularities, nevertheless, with a clever analytical rearrangement, the absence of noncausal singularities is achieved and leads to a causal representation which is more efficient and stable numerically in all the integration domain.
To illustrate the impact of noncausal singularities, we present in Fig.~\ref{fig:N3MLTtIntegrand} the integrand of the dual representation of the N$^3$MLT vacuum diagram ($k_{j,0}=0$) as a function of the two on-shell energies $\qon{12}$ and $\qon{123}$ where the remaining on-shell energies are set at fixed values, $\qon{i}=1$ for $i=1,\dots,5$ and $\qon{234}=2$. The white lines represent the location of the noncausal thresholds which arise due to the denominators $1/(\qon{12}-\qon{i})$, $1/(\qon{123}-\qon{i})$ and $1/(\qon{12}-\qon{123}\pm\qon{i})$. 
Additionally, to clarify the meaning of the noncausal thresholds in Fig.~\ref{fig:N3MLTtIntegrand} in more detail, we study one of the singularities by fixing $\qon{123}$ and scanning over $\qon{12}$. The results of noncausal and causal evaluations of the N$^3$MLT configurations are displayed in Fig.~\ref{fig:N3MLTtestabilidad} in the left and right plots, respectively.  
Similar findings were reported in Ref. \cite{Aguilera-Verdugo:2020kzc}, where explicit causal representations of up to N$^2$MLT complexity were presented.

In return, the LTD representation in \Eq{eq:master} is universal and valid regardless of the internal configuration, 
while the causal representation is specific to the details of the configuration under consideration, e.g., the number 
of propagators in each loop set. The number of terms for a given N$^{k-1}$MLT topology in \Eq{eq:master} 
scales with the number of loops and linearly with the number of propagators per loop set, but the sum
over residues, equivalently over internal propagators, is implicitly accounted in this expression. 
By contrast, the number of terms for a given N$^{k-1}$MLT topology is independent of the number of loops in the causal representation 
but requires to specify additional causal thresholds and additional causal entanglements when more internal propagators are considered.  
In this respect, external momenta attached to interaction vertices that connect different loop sets do 
not alter the number of internal propagators and therefore the complexity of the causal representation. We will exploit 
this feature in the following to simplify the discussion of the causal representation of the N$^4$MLT topology. 
The full causal expressions with external momenta can be deduced from the causal representation of the vacuum configuration
by matching the momentum flows of the entangled thresholds.   

\begin{figure}[t!]
\includegraphics[scale=0.7]{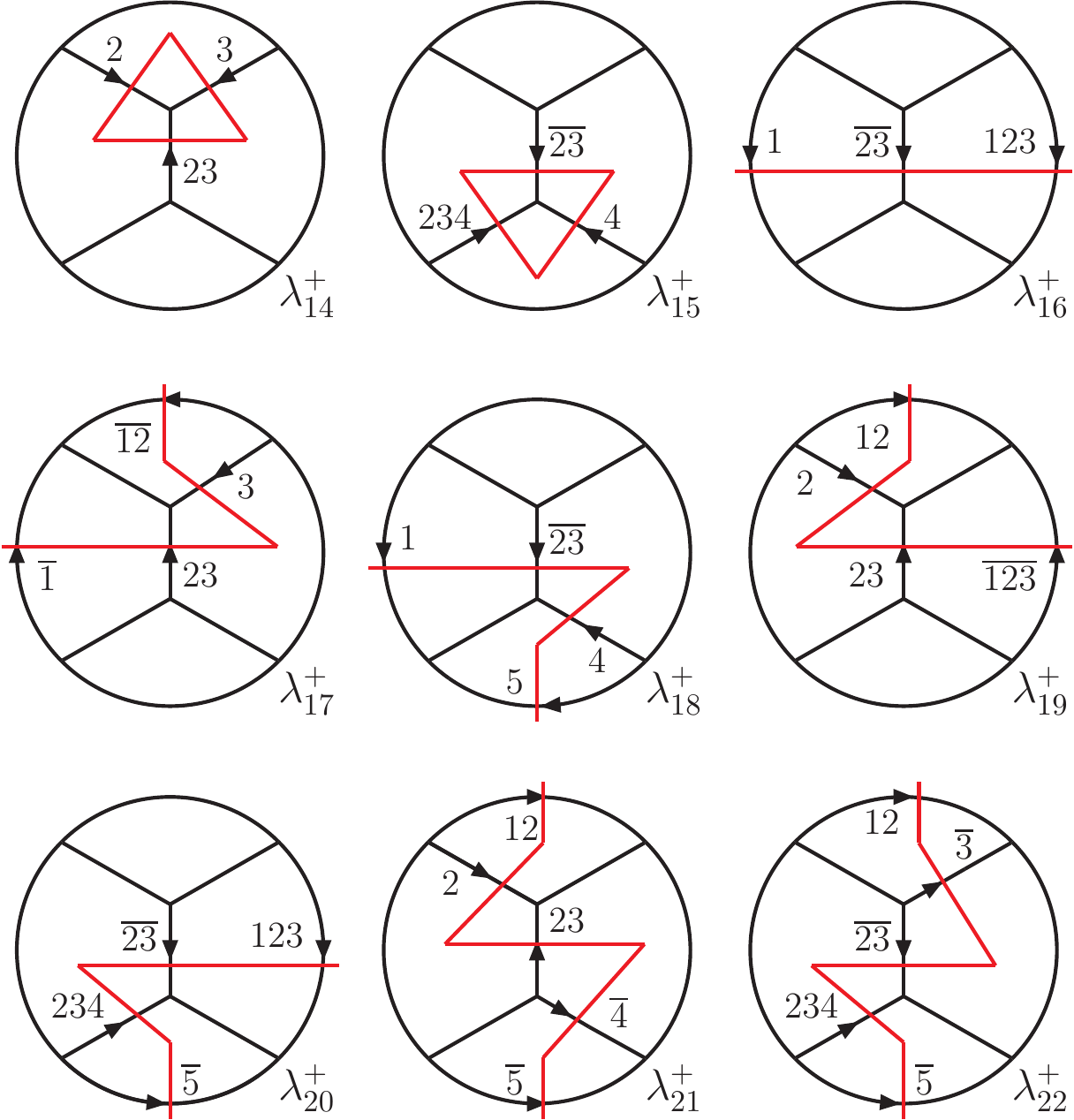}
\centering
\caption{Extra causal configurations of the $t$-channel of the ${\rm N^4MLT}$ topology. 
\label{fig:N4MLTtcausal}}
\end{figure}

\par\bigskip Let us then consider the $t$-channel of the N$^4$MLT universal topology 
\beq
{\cal A}_{{\rm N}^4{\rm MLT}}^{(L)} (1, \ldots, L+4, 23) = \int_{\ell_1, \cdots, \ell_L} \mathcal{N}( \{ \ell_s\}_L,  \{ p_j\}_N) \, G_F(1, \ldots, L+4, 23)~,
\eeq
again with only one propagator per set, and six external particles. 
Its LTD representation is causal after all the nested residues are summed up together, i.e. 
\beq
{\cal A}_{{\rm N}^4{\rm MLT}}^{(L)} (1, \ldots, L+4, 23) = 
\int_{\vec \ell_1, \cdots, \vec \ell_L} \frac{{\cal N}_{{\rm N}^4{\rm MLT}} (\{\qon{s}, k_{j,0}\})}{x_{t, L+5} \left( \prod_{i=1}^{22} \lambda_i^+ \lambda_i^- \right)}~,
\label{eq:N4MLTtiscausal}
\eeq
where $x_{t, L+5} = 2 \qon{23} \, x_{L+4}$.
The LTD representation in \Eq{eq:N4MLTtiscausal} depends on the causal denominators already defined for the N$^3$MLT configuration in 
\Eq{eq:lambdaN3MLT} in addition to nine extra causal denominators that depend on $\qon{23}$ 
(the corresponding configurations are shown in Fig.~\ref{fig:N4MLTtcausal}):
\begin{align}
&  \lambda_{14}^\pm = \qon{(2,3, 23)} \pm k_{23,0}~, 
&&  \lambda_{15}^\pm = \qon{(4, 234,23)} \pm (k_{234}-k_{23})_0~, \nn \\ 
&   \lambda_{16}^\pm = \qon{(1, 123, 23)} \pm (k_{123}-k_{23})_0~,  
&&  \lambda_{17}^\pm = \qon{(1, 3, 12, 23)} \pm (k_{23}-k_{12})_0~, \nn \\
&   \lambda_{18}^\pm = \qon{(1,4, \ldots, L+1, 23)}  \pm (k_{L+1}-k_{23})_0~, \nn \\
&   \lambda_{19}^\pm = \qon{(2, 12, 123, 23)} \pm (k_{12}+k_{23}-k_{123})_0~, \nn \\
& \lambda_{20}^\pm = \qon{(123,234, 5, \ldots, L+1,23)} \pm (k_{123}+k_{234}-k_{L+1}-k_{23})_0~, \nn \\
&   \lambda_{21}^\pm = \qon{(2, 4, \ldots, L+1, 12, 23)} \pm (k_{12}+k_{23}-k_{L+1})_0~, \nn \\
& \lambda_{22}^\pm = \qon{(3, 5, \ldots, L+1, 12, 234, 23)} \pm (k_{234}+k_{12}-k_{23}-k_{L+1})_0~. 
\label{eq:newlambda}
\end{align}
The numerator in \Eq{eq:N4MLTtiscausal} is now a polynomial of degree seventeen for the scalar integral. 
We should consider then all the entangled configurations
with five causal thresholds. For the sake of simplicity, 
we will restrict the analysis to the vacuum configuration, which implies $\lambda_i^+= \lambda_i^-$
and is sufficient to have a clear overview of the causal structure, as explained before. 
Applying the reconstruction algorithm defined in Ref.~\cite{Aguilera-Verdugo:2020kzc}, 
we obtain again a very compact result, and an overall structure similar to \Eq{eq:N3MLTcausal}, 
\bea
&& {\cal A}_{{\rm N}^4{\rm MLT}}^{(L)} (1,\ldots, L+4, 23) = - \int_{\vec \ell_1, \cdots, \vec \ell_L} \frac{2}{x_{t, L+5}}  
\bigg[ {\cal F}^{(L+5)}_{(1,5,6,7,10,11,12,13,14,15,18)}  \nn \\ 
&& \quad + {\cal F}^{(L+5)}_{(2,5,7,8,11,12,13,10,15,14,17)} + {\cal F}^{(L+5)}_{(3,5,8,9,12,13,10,11,15,14,19)} 
+ {\cal F}^{(L+5)}_{(4,5,9,6,13,10,11,12,14,15,20)} \nn \\
&& \quad + {\cal F}^{(L+5)}_{(21,18,19,7,14,11,12,9,13,15,10)} + {\cal F}^{(L+5)}_{(22,17,20,7,15,12,11,9,10,14,13)} \nn \\
&& \quad + {\cal G}^{(L+5)}_{(10,14,6,13,11,12,15,8)}
+ {\cal G}^{(L+5)}_{(10,11,12,13,14,15,17,18)} + {\cal G}^{(L+5)}_{(12,13,10,11,15,14,20,19)} \nn \\
&& \quad + \left( L_{(6,10,11)} + L_{(17,10,14)} + L_{(19,11,14)} \right)
\left( L_{(8,12,13)} + L_{(18,13,15)} + L_{(20,12,15)} \right) L_{16}  \bigg]~, \nn \\
\label{eq:N4MLTtcausal}
\eea 
that is written in terms of permutations of the arguments of the two functions 
\bea
{\cal F}^{(L+5)}_{(1,\ldots, 11)} &=& L_1
\left( L_{(2, 9,10)} + L_{(8, 9,11)} + L_{10} \,L_{11} \right) 
\left( L_{(3, 5,6)} + L_{(4, 6,7)} + L_{5} \,L_{7} \right)~, 
\eea
and
\beq
{\cal G}^{(L+5)}_{(1,\ldots, 8)} = \left( L_{1} + L_{5} \right)  \, L_{(3,4,6)} \, L_{(7,2,8)}~.
\eeq 

Notice that, for example
\bea
&& \left. {\cal F}^{(L+5)}_{(1,5,6,7,10,11,12,13,14,15,18)} \right|_{L_{14}\to 1, (L_{15}, L_{18})\to 0}
= {\cal F}^{(L+4)}_{(1,5,6,7,10,11,12,13)}~,
\eea
thus ensuring the consistency of \Eq{eq:N4MLTtcausal} with \Eq{eq:N3MLTcausal}.

The $s$-channel is obtained just by a clockwise rotation of the $t$-channel and therefore by a permutation of the 
arguments of the causal denominators that are channel specific
\bea
&& \lambda_i^\pm (1, 2, 3, 4, 5\cdots(L+1), 12, 123, 234, 34) \nn \\ 
&& \qquad = \lambda_i^\pm (\overline{5\cdots(L+1)}, 234, 2, 3, 123, \overline{1},12, 4, 23)~,  \qquad i \in [14,22]~.
\eea
This means that in \Eq{eq:newlambda} it is enough to make, for example, replacements similar to 
\bea
&& \lambda_{14}^\pm \to \qon{(3,4,34)} \pm k_{34,0}~. \nn \\
&& \lambda_{18}^\pm \to \qon{(1,12, 234, 34)} \pm (k_{234}-k_{12}-k_{34})_0~,
\eea
to obtain the corresponding causal representation. 

\begin{figure}[t!]
\includegraphics[scale=0.7]{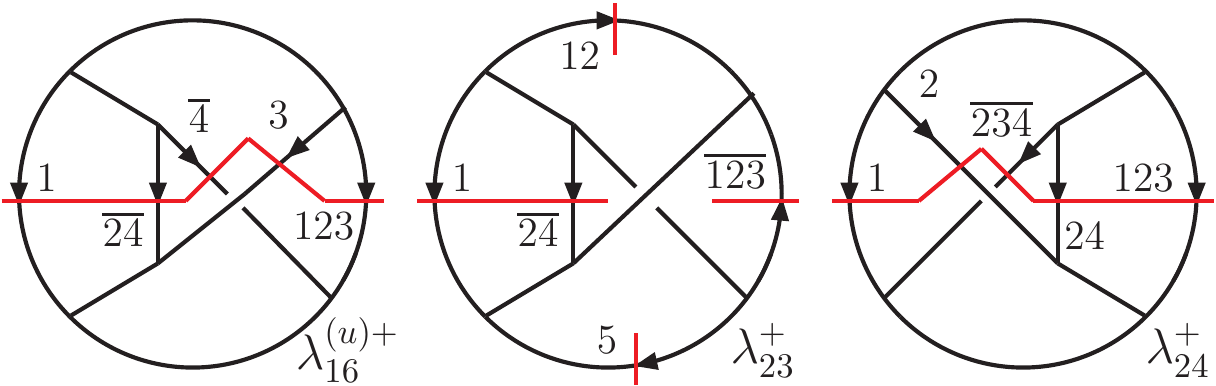}
\centering
\caption{Extra causal configurations of the $u$-channel of the ${\rm N^4MLT}$ topology due to nonplanarity.
\label{fig:N4MLTucausal}}
\end{figure}

\par\bigskip For the $u$-channel, the causal denominators are obtained from the $t$-channel through the substitution
$23\to 24$ and by the exchange $3\leftrightarrow 4$ or $2\leftrightarrow 234$  ($123$ remains invariant):
\bea
\lambda_i^\pm (1, \ldots, L+1, 12, 123, 234, 24) &=& \lambda_i^\pm (3 \leftrightarrow 4, 23\to 24)~,  \qquad i \in [14,15,17,18,22]~, \nn \\
\lambda_i^\pm (1, \ldots, L+1, 12, 123, 234, 24) &=& \lambda_i^\pm (2 \leftrightarrow 234, 23\to 24)~,  \qquad i \in [19,20,21]~.
\eea
There are, however, three new configurations that arise because the $u$-channel is nonplanar. These new configurations
are shown in  Fig.~\ref{fig:N4MLTucausal} and are described by the following causal denominators
\bea
&& \lambda_{16}^{(u)\pm} = \qon{(1, \ldots, 4, 123)} \pm k_{123,0}~, \nn \\
&& \lambda_{23}^\pm = \qon{(1, 5, \ldots, L+1,12,123, 24)} \pm  (k_{L+1}+k_{12} - k_{123}-k_{24})_0~, \nn \\
&& \lambda_{24}^\pm = \qon{(1, 2, 123, 234, 24)} \pm (k_{123}-k_{234}+k_{24})_0~.
\eea
The causal representation of the $u$-channel has a very similar structure to \Eq{eq:N4MLTtcausal}
\bea
&& {\cal A}_{{\rm N}^4{\rm MLT}}^{(L)} (1,\ldots, L+4, 24) = - \int_{\vec \ell_1, \cdots, \vec \ell_L} \frac{2}{x_{u, L+5}}  
\bigg[ {\cal F}^{(L+5)}_{(1,5,6,7,10,11,12,13,14,15,18)}  \nn \\ 
&& \quad + {\cal F}^{(L+5)}_{(2,5,7,8,11,12,13,10,15,14,17)}  + {\cal F}^{(L+5)}_{(3,5,8,9,12,13,10,11,14,15,19)} 
+ {\cal F}^{(L+5)}_{(4,5,9,6,13,10,11,12,15,14,20)}  \nn \\
&& \quad + {\cal F}^{(L+5)}_{(21,18,20,7,14,12,11,9,13,15,10)} + {\cal F}^{(L+5)}_{(22,17,19,7,15,11,12,9,10,14,13)} 
+ {\cal F}^{(L+5)}_{(16,18,17,6,14,10,11,8,13,15,12)} \nn \\
&& \quad + {\cal F}^{(L+5)}_{(24,20,19,6,15,11,10,8,12,14,13)} + {\cal F}^{(L+5)}_{(23,19,20,17,12,14,10,18,15,11,13)} \nn \\
&& \quad + L_{10} \, L_{12} \, L_{15}
 \left( L_{11} \, L_{13} + L_{11} \, L_{14} + L_{13} \, L_{14}
\right) \bigg]~,
\label{eq:N4MLTucausal}
\eea 
where $x_{u, L+5} = 2 \qon{24} \, x_{L+4}$.

\section{Conclusions}

We have analized the multiloop topologies that appear for the first time at four loops and have found a general representation, 
the N$^4$MLT {\it universal topology}, which describes their opening to nondisjoint trees through the loop-tree duality. 
The opening to trees admits a very structured and compact factorized interpretation in terms of convolutions of known subtopologies,
that finally determine the internal causal structure of the entire amplitude. 
The LTD representation presented in this paper is valid in arbitrary coordinate systems and space-time dimensions.

The N$^4$MLT topology is called universal because it unifies in a single expression all the necessary ingredients to 
open any scattering amplitude of up to four loops. Beyond four loops, 
this topology will be embedded in more complex topologies, so that the methodology presented here can 
be used as a guide to achieve a full description of higher orders.

We have verified that the LTD representation of N$^4$MLT is manifestly causal,
namely, that the explicit LTD analytic expression is inherently free of noncausal singularities. 
On the one hand, this supports the applicability and generalization of four-dimensional unsubtraction to higher orders. 
On the other hand, it allows a more efficient numerical evaluation of multiloop scattering amplitudes
than other integrand representations due to the absence of noncausal singularities.
These results extend by one perturbative order the causal analysis of Ref.~\cite{Aguilera-Verdugo:2020kzc}, 
and the interpretation of LTD in terms of entangled causal thresholds. 
In addition, they confirm the all-order conjectures of Ref.~\cite{Verdugo:2020kzh}.
We expect similar conclusions at higher orders, thus leading to a noticeable improvement in the available toolkit for computing highly-precise theoretical predictions.


\section{Acknowledgments}
This work is supported by the Spanish Government
(Agencia Estatal de Investigaci\'on) and ERDF funds from
European Commission (Grant No. FPA2017-84445-P), Generalitat Valenciana (Grant
No. PROMETEO/2017/053),
and the COST Action CA16201 PARTICLEFACE. S. R.-U. 
from CONACYT and Universidad Aut\'onoma de Sinaloa; 
W.J.T. from Juan de la Cierva program (FJCI-2017-32128); and 
R. J. H.-P. from Departament de F\'isica Te\`orica, Universitat 
de Val\`encia, CONACyT through the Project No. A1-S-33202 
(Ciencia B\'asica) and Sistema Nacional de Investigadores.

\bibliographystyle{JHEP}

\begin{thebibliography}{10}

\bibitem{Mangano:2020icy}
M.~Mangano, \emph{{LHC at 10: the physics legacy}},
  \href{http://arxiv.org/abs/2003.05976}{{\tt 2003.05976}}.

\bibitem{Abada:2019lih}
{\scshape FCC} collaboration, A.~Abada et~al., \emph{{FCC Physics
  Opportunities}: {Future Circular Collider Conceptual Design Report Volume
  1}}, \href{http://dx.doi.org/10.1140/epjc/s10052-019-6904-3}{\emph{Eur. Phys.
  J. C} {\bf 79} (2019) 474}.

\bibitem{Abada:2019zxq}
{\scshape FCC} collaboration, A.~Abada et~al., \emph{{FCC-ee: The Lepton
  Collider}: {Future Circular Collider Conceptual Design Report Volume 2}},
  \href{http://dx.doi.org/10.1140/epjst/e2019-900045-4}{\emph{Eur. Phys. J. ST}
  {\bf 228} (2019) 261--623}.

\bibitem{Abada:2019ono}
{\scshape FCC} collaboration, A.~Abada et~al., \emph{{HE-LHC: The High-Energy
  Large Hadron Collider}: {Future Circular Collider Conceptual Design Report
  Volume 4}}, \href{http://dx.doi.org/10.1140/epjst/e2019-900088-6}{\emph{Eur.
  Phys. J. ST} {\bf 228} (2019) 1109--1382}.

\bibitem{Benedikt:2018csr}
{\scshape FCC} collaboration, A.~Abada et~al., \emph{{FCC-hh: The Hadron
  Collider}: {Future Circular Collider Conceptual Design Report Volume 3}},
  \href{http://dx.doi.org/10.1140/epjst/e2019-900087-0}{\emph{Eur. Phys. J. ST}
  {\bf 228} (2019) 755--1107}.

\bibitem{Bambade:2019fyw}
P.~Bambade et~al., \emph{{The International Linear Collider: A Global
  Project}},  \href{http://arxiv.org/abs/1903.01629}{{\tt 1903.01629}}.

\bibitem{Djouadi:2007ik}
{\scshape ILC} collaboration, G.~Aarons et~al., \emph{{International Linear
  Collider Reference Design Report Volume 2: Physics at the ILC}},
  \href{http://arxiv.org/abs/0709.1893}{{\tt 0709.1893}}.

\bibitem{Roloff:2018dqu}
{\scshape CLIC, CLICdp} collaboration, \emph{{The Compact Linear e$^+$e$^-$
  Collider (CLIC): Physics Potential}},
  \href{http://arxiv.org/abs/1812.07986}{{\tt 1812.07986}}.

\bibitem{CEPCStudyGroup:2018ghi}
{\scshape CEPC Study Group} collaboration, M.~Dong et~al., \emph{{CEPC
  Conceptual Design Report: Volume 2 - Physics \& Detector}},
  \href{http://arxiv.org/abs/1811.10545}{{\tt 1811.10545}}.

\bibitem{Anastasiou:2015vya}
C.~Anastasiou, C.~Duhr, F.~Dulat, F.~Herzog and B.~Mistlberger, \emph{{Higgs
  Boson Gluon-Fusion Production in QCD at Three Loops}},
  \href{http://dx.doi.org/10.1103/PhysRevLett.114.212001}{\emph{Phys. Rev.
  Lett.} {\bf 114} (2015) 212001}, [\href{http://arxiv.org/abs/1503.06056}{{\tt
  1503.06056}}].

\bibitem{Mondini:2019gid}
R.~Mondini, M.~Schiavi and C.~Williams, \emph{{N$^{3}$LO predictions for the
  decay of the Higgs boson to bottom quarks}},
  \href{http://dx.doi.org/10.1007/JHEP06(2019)079}{\emph{JHEP} {\bf 06} (2019)
  079}, [\href{http://arxiv.org/abs/1904.08960}{{\tt 1904.08960}}].

\bibitem{Mistlberger:2018etf}
B.~Mistlberger, \emph{{Higgs boson production at hadron colliders at N$^{3}$LO
  in QCD}}, \href{http://dx.doi.org/10.1007/JHEP05(2018)028}{\emph{JHEP} {\bf
  05} (2018) 028}, [\href{http://arxiv.org/abs/1802.00833}{{\tt 1802.00833}}].

\bibitem{Ahmed:2016otz}
T.~Ahmed, M.~Bonvini, M.~Kumar, P.~Mathews, N.~Rana, V.~Ravindran et~al.,
  \emph{{Pseudo-scalar Higgs boson production at N$^3$ LO$_{\text {A}}$ +N$^3$
  LL $'$}}, \href{http://dx.doi.org/10.1140/epjc/s10052-016-4510-1}{\emph{Eur.
  Phys. J. C} {\bf 76} (2016) 663},
  [\href{http://arxiv.org/abs/1606.00837}{{\tt 1606.00837}}].

\bibitem{Bonetti:2017ovy}
M.~Bonetti, K.~Melnikov and L.~Tancredi, \emph{{Three-loop mixed
  QCD-electroweak corrections to Higgs boson gluon fusion}},
  \href{http://dx.doi.org/10.1103/PhysRevD.97.034004}{\emph{Phys. Rev. D} {\bf
  97} (2018) 034004}, [\href{http://arxiv.org/abs/1711.11113}{{\tt
  1711.11113}}].

\bibitem{Dreyer:2016oyx}
F.~A. Dreyer and A.~Karlberg, \emph{{Vector-Boson Fusion Higgs Production at
  Three Loops in QCD}},
  \href{http://dx.doi.org/10.1103/PhysRevLett.117.072001}{\emph{Phys. Rev.
  Lett.} {\bf 117} (2016) 072001}, [\href{http://arxiv.org/abs/1606.00840}{{\tt
  1606.00840}}].

\bibitem{Cieri:2018oms}
L.~Cieri, X.~Chen, T.~Gehrmann, E.~N. Glover and A.~Huss, \emph{{Higgs boson
  production at the LHC using the $q_T$ subtraction formalism at N$^3$LO QCD}},
  \href{http://dx.doi.org/10.1007/JHEP02(2019)096}{\emph{JHEP} {\bf 02} (2019)
  096}, [\href{http://arxiv.org/abs/1807.11501}{{\tt 1807.11501}}].

\bibitem{Cieri:2020ikq}
L.~Cieri, D.~de~Florian, M.~Der and J.~Mazzitelli, \emph{{Mixed QCD$\otimes$QED
  corrections to exclusive Drell Yan production using the $q_T$-subtraction
  method}},  \href{http://arxiv.org/abs/2005.01315}{{\tt 2005.01315}}.

\bibitem{Catani:2008xa}
S.~Catani, T.~Gleisberg, F.~Krauss, G.~Rodrigo and J.-C. Winter, \emph{{From
  loops to trees by-passing Feynman's theorem}},
  \href{http://dx.doi.org/10.1088/1126-6708/2008/09/065}{\emph{JHEP} {\bf 09}
  (2008) 065}, [\href{http://arxiv.org/abs/0804.3170}{{\tt 0804.3170}}].

\bibitem{Bierenbaum:2010cy}
I.~Bierenbaum, S.~Catani, P.~Draggiotis and G.~Rodrigo, \emph{{A Tree-Loop
  Duality Relation at Two Loops and Beyond}},
  \href{http://dx.doi.org/10.1007/JHEP10(2010)073}{\emph{JHEP} {\bf 10} (2010)
  073}, [\href{http://arxiv.org/abs/1007.0194}{{\tt 1007.0194}}].

\bibitem{Bierenbaum:2012th}
I.~Bierenbaum, S.~Buchta, P.~Draggiotis, I.~Malamos and G.~Rodrigo,
  \emph{{Tree-Loop Duality Relation beyond simple poles}},
  \href{http://dx.doi.org/10.1007/JHEP03(2013)025}{\emph{JHEP} {\bf 03} (2013)
  025}, [\href{http://arxiv.org/abs/1211.5048}{{\tt 1211.5048}}].

\bibitem{Tomboulis:2017rvd}
E.~Tomboulis, \emph{{Causality and Unitarity via the Tree-Loop Duality
  Relation}}, \href{http://dx.doi.org/10.1007/JHEP05(2017)148}{\emph{JHEP} {\bf
  05} (2017) 148}, [\href{http://arxiv.org/abs/1701.07052}{{\tt 1701.07052}}].

\bibitem{Runkel:2019yrs}
R.~Runkel, Z.~Sz\H{o}r, J.~P. Vesga and S.~Weinzierl, \emph{{Causality and
  loop-tree duality at higher loops}},
  \href{http://dx.doi.org/10.1103/PhysRevLett.122.111603,
  10.1103/PhysRevLett.123.059902}{\emph{Phys. Rev. Lett.} {\bf 122} (2019)
  111603}, [\href{http://arxiv.org/abs/1902.02135}{{\tt 1902.02135}}].

\bibitem{Capatti:2019ypt}
Z.~Capatti, V.~Hirschi, D.~Kermanschah and B.~Ruijl, \emph{{Loop-Tree Duality
  for Multiloop Numerical Integration}},
  \href{http://dx.doi.org/10.1103/PhysRevLett.123.151602}{\emph{Phys. Rev.
  Lett.} {\bf 123} (2019) 151602}, [\href{http://arxiv.org/abs/1906.06138}{{\tt
  1906.06138}}].

\bibitem{Verdugo:2020kzh}
J.~J. Aguilera-Verdugo, F.~Driencourt-Mangin, R.~J. Hernandez~Pinto,
  J.~Plenter, S.~Ramirez-Uribe, A.~E. Renteria~Olivo et~al., \emph{{Open loop
  amplitudes and causality to all orders and powers from the loop-tree
  duality}},
  \href{http://dx.doi.org/10.1103/PhysRevLett.124.211602}{\emph{Phys. Rev.
  Lett.} {\bf 124} (2020) 211602}, [\href{http://arxiv.org/abs/2001.03564}{{\tt
  2001.03564}}].

\bibitem{Buchta:2014dfa}
S.~Buchta, G.~Chachamis, P.~Draggiotis, I.~Malamos and G.~Rodrigo, \emph{{On
  the singular behaviour of scattering amplitudes in quantum field theory}},
  \href{http://dx.doi.org/10.1007/JHEP11(2014)014}{\emph{JHEP} {\bf 11} (2014)
  014}, [\href{http://arxiv.org/abs/1405.7850}{{\tt 1405.7850}}].

\bibitem{Aguilera-Verdugo:2019kbz}
J.~J. Aguilera-Verdugo, F.~Driencourt-Mangin, J.~Plenter,
  S.~Ram{\'\i}rez-Uribe, G.~Rodrigo, G.~F. Sborlini et~al., \emph{{Causality,
  unitarity thresholds, anomalous thresholds and infrared singularities from
  the loop-tree duality at higher orders}},
  \href{http://dx.doi.org/10.1007/JHEP12(2019)163}{\emph{JHEP} {\bf 12} (2019)
  163}, [\href{http://arxiv.org/abs/1904.08389}{{\tt 1904.08389}}].

\bibitem{Hernandez-Pinto:2015ysa}
R.~J. Hernandez-Pinto, G.~F.~R. Sborlini and G.~Rodrigo, \emph{{Towards gauge
  theories in four dimensions}},
  \href{http://dx.doi.org/10.1007/JHEP02(2016)044}{\emph{JHEP} {\bf 02} (2016)
  044}, [\href{http://arxiv.org/abs/1506.04617}{{\tt 1506.04617}}].

\bibitem{Sborlini:2016gbr}
G.~F.~R. Sborlini, F.~Driencourt-Mangin, R.~Hernandez-Pinto and G.~Rodrigo,
  \emph{{Four-dimensional unsubtraction from the loop-tree duality}},
  \href{http://dx.doi.org/10.1007/JHEP08(2016)160}{\emph{JHEP} {\bf 08} (2016)
  160}, [\href{http://arxiv.org/abs/1604.06699}{{\tt 1604.06699}}].

\bibitem{Sborlini:2016hat}
G.~F.~R. Sborlini, F.~Driencourt-Mangin and G.~Rodrigo, \emph{{Four-dimensional
  unsubtraction with massive particles}},
  \href{http://dx.doi.org/10.1007/JHEP10(2016)162}{\emph{JHEP} {\bf 10} (2016)
  162}, [\href{http://arxiv.org/abs/1608.01584}{{\tt 1608.01584}}].

\bibitem{Driencourt-Mangin:2019sfl}
F.~Driencourt-Mangin, \emph{{Four-dimensional representation of scattering
  amplitudes and physical observables through the application of the Loop-Tree
  Duality theorem}}.
\newblock PhD thesis, U. Valencia (main), 2019.
\newblock \href{http://arxiv.org/abs/1907.12450}{{\tt 1907.12450}}.

\bibitem{Soper:1999xk}
D.~E. Soper, \emph{{Techniques for QCD calculations by numerical integration}},
  \href{http://dx.doi.org/10.1103/PhysRevD.62.014009}{\emph{Phys. Rev. D} {\bf
  62} (2000) 014009}, [\href{http://arxiv.org/abs/hep-ph/9910292}{{\tt
  hep-ph/9910292}}].

\bibitem{Fazio:2014xea}
R.~A. Fazio, P.~Mastrolia, E.~Mirabella and W.~J. Torres~Bobadilla, \emph{{On
  the Four-Dimensional Formulation of Dimensionally Regulated Amplitudes}},
  \href{http://dx.doi.org/10.1140/epjc/s10052-014-3197-4}{\emph{Eur. Phys. J.}
  {\bf C74} (2014) 3197}, [\href{http://arxiv.org/abs/1404.4783}{{\tt
  1404.4783}}].

\bibitem{Freedman:1991tk}
D.~Z. Freedman, K.~Johnson and J.~I. Latorre, \emph{{Differential
  regularization and renormalization: A New method of calculation in quantum
  field theory}},
  \href{http://dx.doi.org/10.1016/0550-3213(92)90240-C}{\emph{Nucl. Phys. B}
  {\bf 371} (1992) 353--414}.

\bibitem{Battistel:1998sz}
O.~Battistel, A.~Mota and M.~Nemes, \emph{{Consistency conditions for 4-D
  regularizations}},
  \href{http://dx.doi.org/10.1142/S0217732398001686}{\emph{Mod. Phys. Lett. A}
  {\bf 13} (1998) 1597--1610}.

\bibitem{Wu:2002xa}
Y.-L. Wu, \emph{{Symmetry principle preserving and infinity free regularization
  and renormalization of quantum field theories and the mass gap}},
  \href{http://dx.doi.org/10.1142/S0217751X03015222}{\emph{Int. J. Mod. Phys.
  A} {\bf 18} (2003) 5363--5420},
  [\href{http://arxiv.org/abs/hep-th/0209021}{{\tt hep-th/0209021}}].

\bibitem{Pittau:2012zd}
R.~Pittau, \emph{{A four-dimensional approach to quantum field theories}},
  \href{http://dx.doi.org/10.1007/JHEP11(2012)151}{\emph{JHEP} {\bf 1211}
  (2012) 151}, [\href{http://arxiv.org/abs/1208.5457}{{\tt 1208.5457}}].

\bibitem{Gnendiger:2017pys}
C.~Gnendiger et~al., \emph{{To ${d}$, or not to ${d}$: recent developments and
  comparisons of regularization schemes}},
  \href{http://dx.doi.org/10.1140/epjc/s10052-017-5023-2}{\emph{Eur. Phys. J.}
  {\bf C77} (2017) 471}, [\href{http://arxiv.org/abs/1705.01827}{{\tt
  1705.01827}}].

\bibitem{Pozzorini:2020hkx}
S.~Pozzorini, H.~Zhang and M.~F. Zoller, \emph{{Rational Terms of UV Origin at
  Two Loops}}, \href{http://dx.doi.org/10.1007/JHEP05(2020)077}{\emph{JHEP}
  {\bf 05} (2020) 077}, [\href{http://arxiv.org/abs/2001.11388}{{\tt
  2001.11388}}].

\bibitem{Cherchiglia:2020iug}
A.~Cherchiglia, D.~Arias-Perdomo, A.~Vieira, M.~Sampaio and B.~Hiller,
  \emph{{Two-loop renormalisation of gauge theories in $4D$ Implicit
  Regularisation: transition rules to dimensional methods}},
  \href{http://arxiv.org/abs/2006.10951}{{\tt 2006.10951}}.

\bibitem{Buchta:2015wna}
S.~Buchta, G.~Chachamis, P.~Draggiotis and G.~Rodrigo, \emph{{Numerical
  implementation of the loop--tree duality method}},
  \href{http://dx.doi.org/10.1140/epjc/s10052-017-4833-6}{\emph{Eur. Phys. J.}
  {\bf C77} (2017) 274}, [\href{http://arxiv.org/abs/1510.00187}{{\tt
  1510.00187}}].

\bibitem{Buchta:2015xda}
S.~Buchta, \emph{{Theoretical foundations and applications of the Loop-Tree
  Duality in Quantum Field Theories}}.
\newblock PhD thesis, Valencia U., 2015.
\newblock \href{http://arxiv.org/abs/1509.07167}{{\tt 1509.07167}}.

\bibitem{Driencourt-Mangin:2019yhu}
F.~Driencourt-Mangin, G.~Rodrigo, G.~F. Sborlini and W.~J. Torres~Bobadilla,
  \emph{{On the interplay between the loop-tree duality and helicity
  amplitudes}},  \href{http://arxiv.org/abs/1911.11125}{{\tt 1911.11125}}.

\bibitem{Capatti:2019edf}
Z.~Capatti, V.~Hirschi, D.~Kermanschah, A.~Pelloni and B.~Ruijl,
  \emph{{Numerical Loop-Tree Duality: contour deformation and subtraction}},
  \href{http://dx.doi.org/10.1007/JHEP04(2020)096}{\emph{JHEP} {\bf 04} (2020)
  096}, [\href{http://arxiv.org/abs/1912.09291}{{\tt 1912.09291}}].

\bibitem{Jurado:2017xut}
J.~L. Jurado, G.~Rodrigo and W.~J. Torres~Bobadilla, \emph{{From Jacobi
  off-shell currents to integral relations}},
  \href{http://dx.doi.org/10.1007/JHEP12(2017)122}{\emph{JHEP} {\bf 12} (2017)
  122}, [\href{http://arxiv.org/abs/1710.11010}{{\tt 1710.11010}}].

\bibitem{Beneke:1997zp}
M.~Beneke and V.~A. Smirnov, \emph{{Asymptotic expansion of Feynman integrals
  near threshold}},
  \href{http://dx.doi.org/10.1016/S0550-3213(98)00138-2}{\emph{Nucl. Phys.}
  {\bf B522} (1998) 321--344}, [\href{http://arxiv.org/abs/hep-ph/9711391}{{\tt
  hep-ph/9711391}}].

\bibitem{Driencourt-Mangin:2017gop}
F.~Driencourt-Mangin, G.~Rodrigo and G.~F. Sborlini, \emph{{Universal dual
  amplitudes and asymptotic expansions for $gg\rightarrow H$ and $H\rightarrow
  \gamma \gamma $ in four dimensions}},
  \href{http://dx.doi.org/10.1140/epjc/s10052-018-5692-5}{\emph{Eur. Phys. J.
  C} {\bf 78} (2018) 231}, [\href{http://arxiv.org/abs/1702.07581}{{\tt
  1702.07581}}].

\bibitem{Plenter:2019jyj}
J.~Plenter, \emph{{Asymptotic Expansions Through the Loop-Tree Duality}},
  \href{http://dx.doi.org/10.5506/APhysPolB.50.1983}{\emph{Acta Phys. Polon. B}
  {\bf 50} (2019) 1983--1992}.

\bibitem{Plenter:2020lop}
J.~Plenter and G.~Rodrigo, \emph{{Asymptotic expansions through the loop-tree
  duality}},  \href{http://arxiv.org/abs/2005.02119}{{\tt 2005.02119}}.

\bibitem{Driencourt-Mangin:2019aix}
F.~Driencourt-Mangin, G.~Rodrigo, G.~F.~R. Sborlini and W.~J. Torres~Bobadilla,
  \emph{{Universal four-dimensional representation of $H \to \gamma \gamma$ at
  two loops through the Loop-Tree Duality}},
  \href{http://dx.doi.org/10.1007/JHEP02(2019)143}{\emph{JHEP} {\bf 02} (2019)
  143}, [\href{http://arxiv.org/abs/1901.09853}{{\tt 1901.09853}}].

\bibitem{Aguilera-Verdugo:2020kzc}
J.~J. Aguilera-Verdugo, R.~J. Hernandez-Pinto, G.~Rodrigo, G.~F. Sborlini and
  W.~J. Torres~Bobadilla, \emph{{Causal representation of multi-loop amplitudes
  within the loop-tree duality}},  \href{http://arxiv.org/abs/2006.11217}{{\tt
  2006.11217}}.

\bibitem{Bollini:1972ui}
C.~G. Bollini and J.~J. Giambiagi, \emph{{Dimensional Renormalization: The
  Number of Dimensions as a Regularizing Parameter}},
  \href{http://dx.doi.org/10.1007/BF02895558}{\emph{Nuovo Cim.} {\bf B12}
  (1972) 20--26}.

\bibitem{tHooft:1972tcz}
G.~'t~Hooft and M.~J.~G. Veltman, \emph{{Regularization and Renormalization of
  Gauge Fields}},
  \href{http://dx.doi.org/10.1016/0550-3213(72)90279-9}{\emph{Nucl. Phys.} {\bf
  B44} (1972) 189--213}.

\bibitem{Aguilera-Verdugo:2020nrp}
J.~J. Aguilera-Verdugo, R.~J. Hernandez-Pinto, G.~Rodrigo, G.~F. Sborlini and
  W.~J. Torres~Bobadilla, \emph{{Mathematical properties of nested residues and
  their application to multi-loop scattering amplitudes}},
  \href{http://arxiv.org/abs/2010.12971}{{\tt 2010.12971}}.

\bibitem{Nogueira:1991ex}
P.~Nogueira, \emph{{Automatic Feynman graph generation}},
  \href{http://dx.doi.org/10.1006/jcph.1993.1074}{\emph{J.Comput.Phys.} {\bf
  105} (1993) 279--289}.

\bibitem{vonManteuffel:2014ixa}
A.~von Manteuffel and R.~M. Schabinger, \emph{{A novel approach to integration
  by parts reduction}},
  \href{http://dx.doi.org/10.1016/j.physletb.2015.03.029}{\emph{Phys. Lett. B}
  {\bf 744} (2015) 101--104}, [\href{http://arxiv.org/abs/1406.4513}{{\tt
  1406.4513}}].

\bibitem{Peraro:2016wsq}
T.~Peraro, \emph{{Scattering amplitudes over finite fields and multivariate
  functional reconstruction}},
  \href{http://dx.doi.org/10.1007/JHEP12(2016)030}{\emph{JHEP} {\bf 12} (2016)
  030}, [\href{http://arxiv.org/abs/1608.01902}{{\tt 1608.01902}}].

\bibitem{Peraro:2019svx}
T.~Peraro, \emph{{FiniteFlow: multivariate functional reconstruction using
  finite fields and dataflow graphs}},
  \href{http://dx.doi.org/10.1007/JHEP07(2019)031}{\emph{JHEP} {\bf 07} (2019)
  031}, [\href{http://arxiv.org/abs/1905.08019}{{\tt 1905.08019}}].

\end{thebibliography}
\providecommand{\href}[2]{#2}\begingroup\raggedright\endgroup

\end{document}